\documentclass[aps, prb, twocolumn, groupedaddress, showpacs]{revtex4}%
\usepackage{amsfonts}
\usepackage{amsmath}
\usepackage{amssymb}
\usepackage{graphicx}
\usepackage{bm}
\usepackage{subfigure}
\usepackage{color}%
\providecommand{\U}[1]{\protect\rule{.1in}{.1in}}
\begin{document}
\title{ Superconducting states and Majorana modes in transition-metal dichalcogenides
under inhomogeneous strain}
\author{Ming-Xun Deng$^{1,2}$}
\author{G. Y. Qi$^{2}$}
\author{W. Luo$^{3}$}
\author{R. Ma$^{4}$}
\author{Rui-Qiang Wang$^{1}$}
\author{R. Shen$^{2,5}$}
\author{L. Sheng$^{2,5}$}
\email{shengli@nju.edu.cn}
\author{D. Y. Xing$^{2,5}$}
\affiliation{$^{1}$Guangdong Provincial Key Laboratory of Quantum Engineering and Quantum
materials, School of Physics and Telecommunication Engineering, South China Normal University, Guangzhou 510006, China}
\affiliation{$^{2}$ National Laboratory of Solid State Microstructures and Department of
Physics, Nanjing University, Nanjing 210093, China}
\affiliation{$^{3}$ School of Science, Jiangxi University of Science and Technology,
Ganzhou 341000, China}
\affiliation{$^{4}$ Jiangsu Key Laboratory for Optoelectronic Detection of Atmosphere and
Ocean, Nanjing University of Information Science and Technology, Nanjing
210044, China }
\affiliation{$^{5}$ Collaborative Innovation Center of Advanced Microstructures, Nanjing
University, Nanjing 210093, China}
\date{\today }

\begin{abstract}
We study the effect of inhomogeneous strain on transition-metal
dichalcogenides with a large intrinsic gap in their spectrum. It is found
that, by tuning the chemical potential, superconductivity can preserve within
the strain-induced discrete pseudo Landau levels (LLs), which introduce
interesting topological properties to these systems. As we show, the
superconductivity for integer fillings is quantum critical, and the quantum
critical coupling strength is determined by the spacing between the two LLs
closest to the Fermi level. For partial fillings, the superconducting gap is
scaled linearly with the coupling strength, and decreases rapidly when the
chemical potential shifts away from the middle of each LL. In the presence of
a Zeeman field, a pair of Majorana modes emerge simultaneously in the two
valleys of strained dichalcogenides. When valley symmetry is further
destroyed, a single Majorana mode can be expected to emerge at the edges of
the strained monolayer dichalcogenides.

\end{abstract}

\maketitle
\affiliation{$^{1}$ National Laboratory of Solid State Microstructures and Department of
Physics, Nanjing University, Nanjing 210093, China}
\affiliation{$^{2}$ Collaborative Innovation Center of Advanced Microstructures, Nanjing
University, Nanjing 210093, China}

%72.80.Ga	Transition-metal compounds
%71.27.+a	Strongly correlated electron systems; heavy fermions
%71.70.Di	Landau levels
%74.90.+n	Other topics in superconductivity (restricted to new topics in section 74)

\section{introduction}

\label{IN}

Since the remarkable discovery of
graphene\cite{Novoselov666,Novoselov:2005aa,Zhang:2005aa}, the study of
physics in atomically thin two dimensional (2D) crystals, which could be of
great potential applications in next-generation nanoelectronic
devices\cite{Novoselov10451,Lee76}, has attracted much attention on both
theoretical and experimental sides\cite{RevModPhys.83.1193,RevModPhys.84.1067}%
. In graphene, the conduction and valence band touch at the corners, referred
to as $K$ and $K^{\prime}$ points, of the 2D hexagonal Brillouin zone. The two
inequivalent points constitute a binary index, termed as the valley index, for
the low energy carriers. In the vicinity of the $K$ ($K^{\prime}$) points, the
low-energy electronic excitations behave as massless Dirac
quasiparticles\cite{Gomes:2012aa}, and the dispersions form a 2D Dirac cone,
whose vertex is called Dirac point. The two valleys are separated far from
each other in the momentum space. For electronic states closed to the Dirac
points, the valley index is expected to be robust against scattering by
perturbations. Therefore, the valley index can be served as a potential
information carrier, the use of which leads to a new concept,
valleytronics\cite{PhysRevLett.97.186404,PhysRevB.77.235406,PhysRevLett.99.236809,Rycerz:2007aa,PhysRevLett.106.156801}%
. When the inversion symmetry is broken, valley Hall effect can
emerge\cite{PhysRevLett.99.236809}, where carriers in different valleys flow
to opposite transverse directions upon application of an electric field.
Moreover, when including the spin-orbit interaction, one can explore spin
physics and spintronics in
graphene\cite{PhysRevLett.106.156801,PhysRevB.74.165310,PhysRevB.75.041401,Avsar:2014aa,PhysRevLett.119.206601}%
.

Interestingly, recent theoretical and experimental studies showed that
intrinsic superconductivity could be induced in graphene under the application
of strain
fields~\cite{Levy544,Gomes:2012aa,PhysRevLett.109.066802,PhysRevLett.108.266801,PhysRevLett.111.046604}%
. The strain introduces pseudo Landau levels (LLs) into graphene, while the
time-reversal (TR) symmetry remains. In the weak coupling regime, the critical
temperature is found to scale linearly with the coupling strength, which is
quite different from the conventional weak-coupling superconductors where the
critical temperature decreases exponentially with the effective
coupling~\cite{PhysRevLett.111.046604}. By modulating the filling factor and
magnitude of strain, one can control the superconducting transition
temperature experimentally, which has profound significance for the
manipulation of quantum states in solid states. Moreover, in the presence of
superconductivity, the system can exhibit exotic topological properties, such
as the emergence of Majorana modes, when the inversion and TR symmetries are
broken spontaneously~\cite{PhysRevB.93.054502}.

Although graphene has many extraordinary physical properties, the inversion
symmetry is preserved and the spin-orbit coupling (SOC) is rather weak in
graphene, which challenges some of its applications in valleytronics and
spintronics. Instead, layered transition-metal
dichalcogenides~\cite{Novoselov10451,Lee76,nl903868w,PhysRevLett.105.136805,Radisavljevic:2011aa,aip_apl99,PhysRevLett.120.106802,PhysRevB.97.075434}%
, with broken inversion symmetry and strong SOC, represent an alternative
class of 2D materials~\cite{Novoselov10451}, which can provide excellent
platforms towards the integration of valleytronics and
spintronics~\cite{Wang:2012aa,PhysRevLett.108.196802}. For example, monolayer
\textrm{MoS}$_{2}$ has similar hexagonal lattice structure as
graphene~\cite{PhysRevLett.108.196802}, but the inversion symmetry is broken
explicitly and the SOC is much stronger in \textrm{MoS}$_{2}$, which makes it
promising candidate materials for valleytronics and
spintronics~\cite{acs.nanolett.7b02364,PhysRevLett.116.046803,PhysRevB.95.205302}%
. In fact, the physics in monolayers of group-VI dichalcogenides $MX_{2}$,
with $M=$\textrm{Mo} and $X=$\textrm{S},\textrm{ Se}, is essentially the same,
all of which are identified as direct-band-gap
semiconductors~\cite{PhysRevB.84.153402}. It is of importance to understand
theoretically how the appearance of the intrinsic band gap and SOC influence
the superconductivity and topological properties of these emergent 2D
materials under strain fields.

In this paper, we investigate the superconductivity and topological properties
of the electrons in strained transition-metal dichalcogenides. Taking
\textrm{MoS}$_{2}$ as an example, we generalize the superconductivity theory
from gapless graphene~\cite{PhysRevLett.111.046604} to strained
dichalcogenides with intrinsic band gaps in their spectrum. We find that the
superconductivity can preserve within the discrete pseudo LLs for these gapped
systems, and the resulting topological phenomena are very interesting. In the
presence of a finite energy gap, the chemical potential plays a very important
role in the occurrence of superconductivity, even for the $n=0$ LL with $n$ as
the LL index. At half fillings, the chemical potential sitting at the middle
of the LLs, the superconductivity gap is maximized. However, at integer
fillings, the emergence of superconductivity requires a minimal quantum
critical coupling, whose strength is determined by the spacing between the LLs
closest to the Fermi level. Below the quantum critical coupling strength, the
superconductivity is fully suppressed. Interestingly, in the presence of a
Zeeman field, a pair of Majorana modes emerge simultaneously in the two
valleys of strained dichalcogenides. A single Majorana mode can be expected to
emerge at the edges of the sample, if the valley symmetry is further destroyed.

The rest of this paper is organized as follows. In the next section, we
introduce the model Hamiltonian and method. The superconductivity in strained
\textrm{MoS}$_{2}$ is discussed in Sec.\ \ref{SS}, and its topological
properties are analyzed in Sec.\ \ref{TIPD}. The final section contains a summary.

\section{Model Hamiltonian and Method}

\label{MHM}

As demonstrated in Refs.\ \cite{PhysRevLett.108.196802,PhysRevB.84.153402},
the underlying physics is the same for monolayers of group-VI dichalcogenides,
such that we can take one of them, i.e., \textrm{MoS}$_{2}$, as an example.
The low-energy electronic excitation in strained monolayer \textrm{MoS}$_{2}$
can be described by the
Hamiltonian~\cite{PhysRevLett.108.266801,PhysRevLett.111.046604,PhysRevLett.108.196802}
\begin{equation}
H=\int d\mathbf{x}\sum_{\xi}\psi_{\xi}^{\dag}(\mathbf{x})H_{\mathbf{p},\xi
}\psi_{\xi}(\mathbf{x}),
\end{equation}
where $\psi_{\xi}(\mathbf{x})=(c_{A,\xi\uparrow},c_{B,\xi\uparrow}%
,c_{A,\xi\downarrow},c_{B,\xi\downarrow})^{T}$, $c_{A(B),\xi\sigma}$ are
electron annihilation operators, and
\begin{equation}
H_{\mathbf{p,}\xi}=\upsilon_{\mathrm{F}}\vec{\Pi}^{\xi}\cdot\vec{\sigma}_{\xi
}+\frac{\Delta_{0}}{2}\sigma_{z}-(\lambda_{\mathrm{so}}\xi\frac{\sigma_{z}%
-1}{2}-m_{\mathrm{z}})s_{z}-\mu\label{Ham0}%
\end{equation}
with $\sigma_{i}$ and $s_{i}$ being the Pauli matrices for sublattice and
spin, respectively, and $\xi=\pm$ representing the valley index. Here,
$\upsilon_{\mathrm{F}}=at/\hbar$ is the Fermi velocity with $a$ and $t$ the
lattice constant and electron hopping integral. $\Delta_{0}$ is the intrinsic
energy gap due to the broken inversion symmetry, $2\lambda_{\mathrm{so}}$ is
the spin splitting at the valence band top due to the spin-orbit coupling,
$m_{\mathrm{z}}$ is the Zeeman field, and $\mu$ is the chemical potential. The
valley-dependent gauge covariant momentum operator $\vec{\Pi}_{x,y}^{\xi}%
=\hat{p}_{x,y}+\xi eA_{x,y}$ is modified by the pseudo-vector-potential
$\mathbf{A}=(\delta t_{x},\delta t_{y})/e\upsilon_{\mathrm{F}}$ generated by
the strain and $\vec{\sigma}_{\xi}=(\xi\sigma_{x},\sigma_{y})$ is a vector of
the Pauli matrices. In the presence of an effective attractive potential $U$,
which stabilizes the superconducting state, the Bogoliubov-de Gennes (BdG)
Hamiltonian is given by $H_{\mathrm{BdG}}=\frac{1}{2}\int d\mathbf{x}\sum
_{\xi}\Psi_{\xi}^{\dag}(\mathbf{x})\mathcal{H}_{\mathrm{BdG}}\Psi_{\xi
}(\mathbf{x})$, where
\begin{equation}
\mathcal{H}_{\mathrm{BdG}}=\left(
\begin{array}
[c]{cc}%
H_{\mathbf{p},\xi}+m_{\mathrm{z}}s_{z} & \hat{\Delta}_{4}\\
\hat{\Delta}_{4}^{\dag} & -\mathcal{T}H_{-\mathbf{p},-\xi}\mathcal{T}%
^{-1}-m_{\mathrm{z}}s_{z}%
\end{array}
\right)  \label{HBdG}%
\end{equation}
and $\Psi_{\xi}^{\dag}(\mathbf{x})=(\psi_{\xi}^{\dag},is_{y}\psi_{-\xi})$,
with $\hat{\Delta}_{4}=\Delta s_{z}\otimes\sigma_{0}$, $\mathcal{T}%
=is_{y}\mathcal{K}$, and $\mathcal{K}$ denoting complex conjugation. In a
proper basis order, the BdG Hamiltonian can be rewritten in the block diagonal
form as%
\begin{equation}
\widetilde{\mathcal{H}}_{\mathrm{BdG}}=\left(
\begin{array}
[c]{cccc}%
h_{\mathbf{p},\xi\uparrow} & \hat{\Delta}_{2} & 0 & 0\\
\hat{\Delta}_{2}^{\dag} & -h_{-\mathbf{p},-\xi\downarrow}^{\ast} & 0 & 0\\
0 & 0 & h_{\mathbf{p},\xi\downarrow} & \hat{\Delta}_{2}\\
0 & 0 & \hat{\Delta}_{2}^{\dag} & -h_{-\mathbf{p},-\xi\uparrow}^{\ast}%
\end{array}
\right)  \label{HBdGb}%
\end{equation}
with $h_{\mathbf{p},\xi\sigma}=\upsilon_{\mathrm{F}}\vec{\Pi}^{\xi}\cdot
\vec{\sigma}_{\xi}+\lambda_{\xi\sigma}\sigma_{z}-\mu_{\xi\sigma}+\sigma
m_{\mathrm{z}}$, where $\lambda_{\xi\sigma}=\frac{\Delta_{0}+\sigma\xi
\lambda_{\mathrm{so}}}{2}$ and $\mu_{\xi\sigma}=\mu+\frac{\sigma\xi
\lambda_{\mathrm{so}}}{2}$.

In the absence of the Zeeman field, the spin index $\sigma=\uparrow
,\downarrow$ is locked to the valley index $\xi$, such that $h_{\mathbf{p}%
,\xi\sigma}=h_{\overline{\mathbf{p}},\overline{\xi}\overline{\sigma}}^{\ast}$,
where $\overline{\sigma}\equiv-\sigma$. As it shows, the strain-induced
pseudomagnetic field does not break the TR symmetry for the system, i.e.,
$H_{-\mathbf{p},-\xi}=\mathcal{T}H_{\mathbf{p},\xi}\mathcal{T}^{-1}$, which is
essentially different from a conventional magnetic field. Therefore,
TR-symmetric states can pair up by the effective attractive interaction, which
favors the formation of Cooper pairs. The pairing matrix $\hat{\Delta}%
_{2}=\Delta\sigma_{0}$ in Eq.\ (\ref{HBdGb}), describing the formation of
Cooper pairs, can be determined self-consistently by $\Delta^{2}=$
$U\mathrm{tr}\langle\psi_{k,\xi\sigma}|\hat{\Delta}_{2}|\psi_{-k,\overline
{\xi}\overline{\sigma}}\rangle$, where $\psi_{k,\xi\sigma}$ is the
two-component spinor for $h_{\mathbf{p},\xi\sigma}$. In the Landau gauge
$\mathbf{A}=(-By,0)$, with $B$ as the pseudomagnetic field, the spinor for
$h_{\mathbf{p},\xi\sigma}$ takes the following form%
\begin{equation}
\psi_{k,\xi\sigma}^{(n)}=\frac{e^{ikx}}{\sqrt{2}}\left(
\begin{array}
[c]{c}%
s_{n}\alpha_{\xi\sigma,+}^{(n)}\phi_{|n|-1}(\zeta)\\
\alpha_{\xi\sigma,-}^{(n)}\phi_{|n|}(\zeta)
\end{array}
\right)  ,
\end{equation}
when $\mu_{\xi\sigma}=m_{\mathrm{z}}=0$. Here, $s_{n}\equiv\mathrm{sgn}(n)$,
$\phi_{|n|}(\zeta)$ is the harmonic wavefunction, and
\begin{equation}
\alpha_{\xi\sigma,\pm}^{(n)}=\sqrt{1\pm\lambda_{\xi\sigma}/\Omega_{\xi\sigma
}^{n}}%
\end{equation}
with $\Omega_{\xi\sigma}^{n}=s_{n}\sqrt{2|n|(\hbar\omega_{c})^{2}+\lambda
_{\xi\sigma}^{2}}-\lambda_{\xi\sigma}\delta_{n,0}$ and $\zeta=\xi
kl_{B}-y/l_{B}$. The cyclotron frequency is defined as $\omega_{c}%
=\upsilon_{\mathrm{F}}/l_{B}$, in which $l_{B}=\sqrt{\hbar/eB}$ denotes the
magnetic length.

\section{Superconducting States}

\label{SS}

Since the Zeeman field is irrelevant to the emergence of superconductivity, we
would omit it in the discussions of the properties of superconducting states,
for simplicity, and it will be considered later in the discussions of the
topological properties in Sec.\ \ref{TIPD}. By projecting
$\widetilde{\mathcal{H}}_{\mathrm{BdG}}$ into the Hilbert space $\{\Phi
^{(n)}\}$, with $\Phi^{(n)}=$ $\left(  \psi_{k,\xi\uparrow}^{(n)}%
,\psi_{-k,\overline{\xi}\downarrow}^{(n)\ast},\psi_{k,\xi\downarrow}%
^{(n)},\psi_{-k,\overline{\xi}\uparrow}^{(n)\ast}\right)  $, the BdG
Hamiltonian can be written in the diagonal form as%
\begin{equation}
\widetilde{\mathcal{H}}_{\mathrm{BdG}}^{(m,n)}=\langle\Phi^{(m)}%
|\widetilde{\mathcal{H}}_{\mathrm{BdG}}|\Phi^{(n)}\rangle=\left(
\begin{array}
[c]{cc}%
\mathcal{H}_{+} & 0\\
0 & \mathcal{H}_{-}%
\end{array}
\right)  \delta_{m,n}\ , \label{hbdgmn}%
\end{equation}
where $\mathcal{H}_{\chi}=(\Omega_{\chi}^{n}-\mu_{\chi})\tau_{z}+\Delta
\tau_{x}$, with $\chi=\pm$ labeling the product of the valley and spin indices
($\xi\times\sigma$) and $\tau_{i}$ being the Pauli matrix for Bogoliubov
isospin in the Nambu space. Some details of derivation are given in the
appendix\ref{append}, e.g., Eq.\ (\ref{eq_BdG}). In the presence of finite
energy gap $\Delta_{0}$ and spin-orbit coupling $\lambda_{\mathrm{so}}$, the
$n=0$ LL $\Omega_{\xi\sigma}^{0}-\mu_{\xi\sigma}$ in strained \textrm{MoS}%
$_{2}$ shifts away from zero energy, which is different from the case in
gapless graphene~\cite{PhysRevLett.111.046604}. As a result, for $\mu=0$, to
preserve the superconductivity in the discrete spectrum of LLs, a finite
coupling is needed to overcome the energy gap between the electron and hole
LLs, even for the $n=0$ LL. With tuning the chemical potential, the electron
and hole LLs could approach each other and encounter at the Fermi level, such
that the gap between the electron and hole LLs would decrease, which reduces
the critical coupling strength. Therefore, in the presence of an intrinsic
gap, the chemical potential plays a very important role in the occurrence of superconductivity.

Due to Pauli blocking, the electronic states are incompressible, so that the
chemical potential would exhibit a discontinuous behavior with changing the
pseudomagnetic field and filling factor. As shown in Eq.\ (\ref{hbdgmn}),
$\mathcal{H}_{\pm}$ are block diagonal and there is no particle exchange
between the two subspaces. Consequently, we can concentrate on one subspace,
say $\mathcal{H}_{\chi}$, and the conclusions can be generalized to the other
subspace straightforwardly. Following Ref.\ \cite{PhysRevLett.111.046604}, we
calculate the chemical potential by fixing the number of particles
$\mathcal{N}_{\chi}$, which can be determined by the fluctuation-dissipation
theorem~\cite{1367-2630-18-9-093040}%
\begin{equation}
\mathcal{N}_{\chi}=-gN_{\phi}\sum_{n}\int_{-\infty}^{\infty}d\omega
\operatorname{Im}[G_{\chi,11}^{r}(\omega)/\pi]f(\omega)\ , \label{FDT_N}%
\end{equation}
where $g=2$ is the valley$\times$spin degeneracy and $N_{\phi}=A/(2\pi
l_{B}^{2})$ is degeneracy of the LL originating from the summation over $k$,
with $A=L_{x}L_{y}$ as area of the sample.\ $f(\omega)=1/(1+e^{\omega/k_{B}%
T})$ is the Fermi-Dirac distribution function and the retarded Green's
function
\begin{align}
G_{\chi}^{r}(\omega)  &  =\frac{1}{\omega+i0^{+}-\mathcal{H}_{\chi}%
}\nonumber\\
&  =\frac{1}{2}\sum_{\eta=\pm}\frac{1}{\omega^{+}-\eta E_{n}^{\chi}}\left[
\tau_{0}+\frac{1}{\eta E_{n}^{\chi}}\left(
\begin{array}
[c]{cc}%
\varepsilon_{n,\chi} & \Delta\\
\Delta & -\varepsilon_{n,\chi}%
\end{array}
\right)  \right]
\end{align}
is a $2\times2$ matrix in the Nambu space, with $E_{n}^{\chi}=\sqrt
{\varepsilon_{n,\chi}^{2}+\Delta^{2}}$ and $\varepsilon_{n,\chi}=\Omega_{\chi
}^{n}-\mu_{\chi}$, where we adopted the abbreviation $\omega^{+}=\omega
+i0^{+}$. The subscripts of the retarded Green's function, e.g., $G_{\chi
,11}^{r}(\omega)$ and $G_{\chi,12}^{r}(\omega)$, stand for the matrix elements
of $G_{\chi}^{r}(\omega)$. If the deep-energy states $n<n_{\mathrm{F}}$ are
fully occupied, with $n_{\mathrm{F}}$ labeling the highest occupied LL, we can
derive%
\begin{equation}
\mathcal{N}_{\chi}=\frac{gN_{\phi}}{2}\sum_{n=-n_{D}}^{n_{\mathrm{F}}}\left(
1-\frac{\varepsilon_{n,\chi}\tanh[E_{n}^{\chi}/(2k_{B}T)]}{E_{n}^{\chi}%
}\right)  \ , \label{slef_Nx}%
\end{equation}
where $n_{D}$ is an ultravialet cutoff related to the bandwidth $D$. More
details are presented in the appendix\ref{append}. \begin{figure}[ptb]
\centering
%Requires \usepackage{graphicx}
\includegraphics[width=\linewidth]{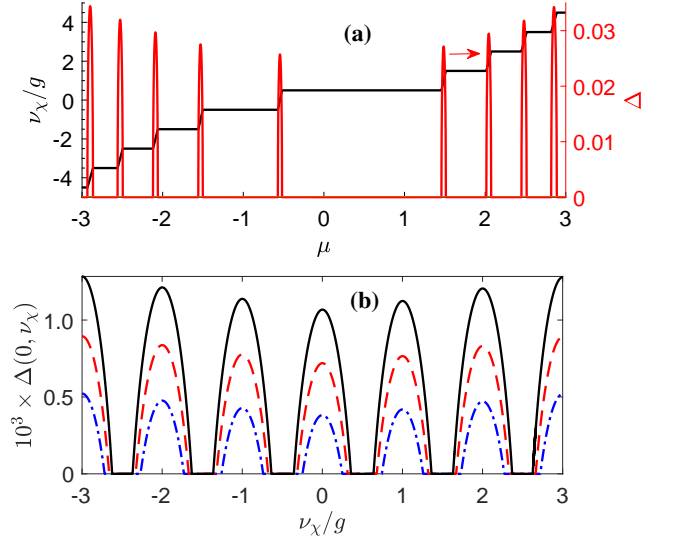} \caption{(a) The filling factor
$\nu_{\chi}/g$ (left) and zero temperature gap $\Delta$ (right) as functions
of the chemical potential $\mu$, for $x=0.035$. (b) The zero temperature gap
$\Delta(0,\nu_{\chi})$ versus the filling factor $\nu_{\chi}/g$ for $x=0.010$,
$0.011$ and $0.012$, from bottom up. Other parameters are set to be $\chi=+$,
$\hbar\omega_{c}=\Delta_{0}$, $2\lambda_{\mathrm{so}}/\Delta_{0}=0.15/1.66$
and $\Delta_{0}=1$ ( $1.66$\textrm{eV }for\textrm{ MoS}$_{2}$%
\cite{PhysRevLett.111.046604} ).}%
\label{FigNG}%
\end{figure}

In the low temperature and weak coupling limit $k_{B}T,\Delta\ll\hbar
\omega_{c}$, we can further reduce Eq.\ (\ref{slef_Nx}) to
\begin{equation}
2(\nu_{\chi}/g-n_{\mathrm{F}})=-\frac{\varepsilon_{n_{\mathrm{F}},\chi}%
\tanh[E_{n_{\mathrm{F}}}^{\chi}/(2k_{B}T)]}{E_{n_{\mathrm{F}}}^{\chi}}
\label{filfactor}%
\end{equation}
with $\nu_{\chi}=N_{\chi}/N_{\phi}-g(n_{D}+1/2)$ being the filling factor. For
$k_{B}T\ll\Delta$, the chemical potential can be approximated to be%
\begin{equation}
\mu_{\chi}(T,\nu_{\chi})=\Omega_{\chi}^{n_{\mathrm{F}}}+\frac{2\Delta
(T,\nu_{\chi})(\nu_{\chi}/g-n_{\mathrm{F}})}{\sqrt{1-4(\nu_{\chi
}/g-n_{\mathrm{F}})^{2}}}. \label{zero_mu}%
\end{equation}
As can be seen, at half fillings $\nu_{\chi}/g=n_{\mathrm{F}}$, the low
temperature chemical potential, pinned to the $n_{\mathrm{F}}$-th LL for
relative small $\Delta(T,\nu_{\chi})$, is robust against the strain. At
integer fillings $\nu_{\chi}/g=n_{\mathrm{F}}\pm1/2$, however,
Eq.\ (\ref{zero_mu}) predicts an unphysical diverging chemical potential if
$\Delta(T,\nu_{\chi})\neq0$. This implies that the superconductivity must be
fully suppressed for integer filling factors, i.e., $\Delta(T,\nu_{\chi})=0$
for $\nu_{\chi}/g=n_{\mathrm{F}}\pm1/2$.

As analyzed, the superconducting gap and chemical potential are interactive,
so that, to determine the superconductivity, the chemical potential must be
accounted self-consistently into the equation of the superconducting gap. gap
equation, as defined in Eq. (\ref{eq_delta}), is obtained, self-consistently,
by%
\begin{equation}
\Delta(T,\nu_{\chi})=-Ug\overline{N}_{\phi}\sum_{n}\int_{-\infty}^{\infty
}d\omega\operatorname{Im}[G_{\chi,12}^{r}(\omega)/\pi]f(\omega) \label{FDT_G}%
\end{equation}
with $\overline{N}_{\phi}=N_{\phi}/A$ as the number of flux quanta per unit
area. Substituting the retarded Green's function, Eq.\ (\ref{eq_retard}), into
Eq. (\ref{FDT_G}) leads to%
\begin{equation}
1=-(U/2)g\overline{N}_{\phi}\sum_{n}\tanh[E_{n}^{\chi}/(2k_{B}T)]/E_{n}^{\chi
}\ . \label{self_gap}%
\end{equation}
Combining Eqs. (\ref{zero_mu}) and (\ref{self_gap}), we obtain for the zero
temperature gap, in the weak coupling regime, as
\begin{equation}
\Delta(0,\nu_{\chi})=\frac{\hbar\upsilon_{\mathrm{F}}\sqrt{1-4(\nu_{\chi
}/g-n_{\mathrm{F}})^{2}}}{2l_{B}[1-\gamma_{n_{\mathrm{F}}}^{(1)}x]}x\ ,
\label{zero_gap}%
\end{equation}
where%
\begin{equation}
\gamma_{n_{\mathrm{F}}}^{(k)}=\sum_{n<n_{\mathrm{F}}}\frac{1}{2|(\varepsilon
_{\chi,n_{\mathrm{F}}}-\varepsilon_{\chi,n})/\hbar\omega_{c}|^{k}}%
\end{equation}
is a constant, and $x=|U|g\overline{N}_{\phi}/(\hbar\omega_{c})$ is a
dimensionless coupling strength of the attractive interaction. As shown by
Eq.\ (\ref{zero_gap}), $\Delta(0,\nu_{\chi})=0$ if $\nu_{\chi}/g=n_{\mathrm{F}%
}\pm1/2$, which confirms the inference that the superconductivity is fully
suppressed for integer filling factors. By substituting Eq.\ (\ref{zero_gap})
into Eq. (\ref{zero_mu}), the zero temperature chemical potential is obtained
as
\begin{equation}
\mu_{\chi}(0,\nu_{\chi})=\Omega_{\chi}^{n_{\mathrm{F}}}+\frac{\hbar
\upsilon_{\mathrm{F}}(\nu_{\chi}/g-n_{\mathrm{F}})}{l_{B}[1-\gamma
_{n_{\mathrm{F}}}^{(1)}x]}x. \label{zero_nv}%
\end{equation}
As it shows, the zero temperature chemical potential at half fillings is
robust to the strain for relative small $x$. At half fillings, the electron
and hole density of states (DOSs) at the Fermi level are maximal, which is
optimum for the emergence of superconductivity. Consequently, the
superconducting gap reaches its maximum for half filling factors, as shown by
Eq.\ (\ref{zero_gap}). Interestingly, the chemical potential, away from the
half fillings $|\nu_{\chi}/g-n_{\mathrm{F}}|>0$, is linearly scaled with the
coupling strength $x$. With $|\nu_{\chi}/g-n_{\mathrm{F}}|$ increasing from
$0$, $\mu_{\chi}(0,\nu_{\chi})$ will shift away from the $n_{\mathrm{F}}$-th
LL and the DOSs at the Fermi level decrease rapidly, which is unfavorable for
the formation of Cooper pairs. As a result, $\Delta(0,\nu_{\chi})$ diminishes
rapidly as the chemical potential shifts away from half fillings. Exactly at
integer fillings, the superconductivity, as indicated by Eq.\ (\ref{zero_gap}%
), is fully suppressed, i.e., $\Delta(0,\nu_{\chi})=0$. As a consequence, the
electronic states, in the weak interaction regime, can not form the
superconducting condensation at integer fillings. \begin{figure}[ptb]
\centering
%Requires \usepackage{graphicx}
\includegraphics[width=\linewidth]{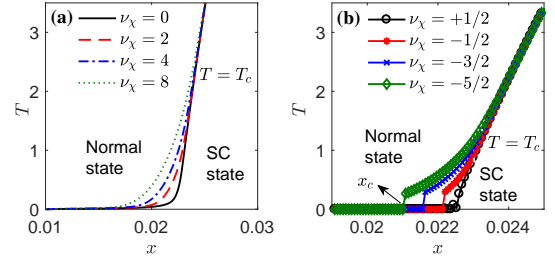} \caption{The phase diagram of
superconducting transition for strained monolayer \textrm{MoS}$_{2}$ for (a)
half fillings and (b) integer fillings. Other parameters are the same as Fig.
\ref{FigNG}.}%
\label{FigTc}%
\end{figure}

The above analysis can be easily verified by the numerical results displayed
in Fig.\ \ref{FigNG}, which is calculated by Eqs.\ (\ref{FDT_N}) and
(\ref{FDT_G}) self-consistently. As shown by the dark curve in
Fig.\ \ref{FigNG}(a), for a fixed $x$, the zero temperature chemical potential
depends linearly on factor $\nu_{\chi}/g-n\in\lbrack-1/2,1/2)$, which
characterizes the width of peaks of the zero temperature gap around half
fillings, as shown by the red curve in Fig.\ \ref{FigNG}(a). The numerical
results are consistent with the analytical ones given by Eq.\ (\ref{zero_nv}).
As shown in Fig.\ \ref{FigNG}(b), for small coupling strength $x\rightarrow0$,
the zero temperature gap peaks around half fillings $\nu_{\chi}/g=n$. The
peaks are separated by zero plateaus in the vicinities of integer fillings
$\nu_{\chi}/g=n\pm1/2$, where formation of Cooper pairs is suppressed. With
increasing the coupling strength, the zero plateaus reduce in width, and
meanwhile, the peaks increase in height rapidly. If the coupling is strong
enough $\Delta\rightarrow\hbar\omega_{c}$, transitions between different LLs
become allowable, such that the zero plateaus could be fully filled by the
peaks, leading to the emergence of superconductivity for integer fillings.
Therefore, there exists a quantum critical coupling strength for the emergence
of superconductivity at integer fillings.

The quantum critical coupling strength can be derived from the expression
(\ref{zero_gap}) for the zero temperature gap. For integer fillings, at which
the chemical potential sits halfway between the two LLs nearest to the Fermi
level, i.e., $\varepsilon_{n_{\mathrm{F}}-1,\chi}=-\varepsilon_{n_{\mathrm{F}%
},\chi}$, the zero temperature energy gap is derived to be
\begin{equation}
\Delta(0,\nu_{\chi}^{\mathrm{I}})=\hbar(\upsilon_{\mathrm{F}}/l_{B}%
)|\Gamma_{\chi}(n_{\mathrm{F}})|\sqrt{(x/x_{c})^{2}-1}%
\end{equation}
where $\nu_{\chi}^{\mathrm{I}}=g(n_{\mathrm{F}}-1/2)$,
\begin{equation}
x_{c}=\frac{1}{\gamma_{n_{\mathrm{F}}-1}^{(1)}+|\Gamma_{\chi}(n_{\mathrm{F}%
})|^{-1}} \label{CriXc}%
\end{equation}
is the critical coupling strength, and
\begin{equation}
\Gamma_{\chi}(n_{\mathrm{F}})=\frac{\varepsilon_{n_{\mathrm{F}},\chi
}-\varepsilon_{n_{\mathrm{F}}-1,\chi}}{2\hbar\omega_{c}}\ .
\end{equation}
As can be seen, the critical coupling strength is determined by the spacing
between the LLs nearest to the Fermi level. For $\Delta_{0}=\lambda
_{\mathrm{so}}=0$, corresponding to case of a graphene sheet, the spacing of
the LLs
\begin{equation}
|\varepsilon_{n,\chi}-\varepsilon_{n-1,\chi}|=\frac{2(\hbar\omega_{c})^{2}%
}{|\Omega_{\chi}^{n}+\Omega_{\chi}^{n-1}|}%
\end{equation}
distributes symmetrically with respect to the $n=0$ LL, and decreases
monotonously with increasing $|n|$. However, in the presence of finite
$\Delta_{0}$ and $\lambda_{\mathrm{so}}$, corresponding to the \textrm{MoS}%
$_{2}$ monolayer, the $0$-th LL $\varepsilon_{0,\chi}$ shifts away from the
symmetry point $\mu=0$, which results in the asymmetric spacings between the
$n=0$ and $n=\pm1$ LLs, i.e., $|\varepsilon_{1,\chi}-\varepsilon_{0,\chi}|>$
$|\varepsilon_{0,\chi}-\varepsilon_{-1,\chi}|$, as demonstrated in
Fig.\ \ref{FigNG}(a). As a result, the critical coupling strength for
$\nu_{\chi}/g=1/2$ is the largest. In other words, as $x$ increases from $0$,
the zero plateaus in Fig.\ \ref{FigNG}(b) will vanish first for the higher
LLs, later for $\nu_{\chi}/g=-1/2$, and last for $\nu_{\chi}/g=1/2$.

For finite temperatures, there exists a critical temperature $T_{c}$, above
which the superconductivity vanishes, i.e., $\Delta(T\geq T_{c},\nu_{\chi}%
)=0$. In Fig.\ \ref{FigTc}, we plot the phase diagram of superconducting
transition for strained monolayer \textrm{MoS}$_{2}$ in the $T-x$ parameter
space. In the critical regime $T\rightarrow T_{c}\gg\Delta$, by using the
Poisson sum formula%
\begin{equation}
f(\xi)=k_{B}T\sum_{m=-\infty}^{\infty}\frac{e^{i\omega_{m}\delta}}{i\omega
_{m}-\xi}%
\end{equation}
with $\omega_{m}=(2m+1)\pi k_{B}T$ and $\delta\rightarrow0^{+}$, we can
convert the gap equation to be
\begin{equation}
1=xk_{B}T\sum_{n}\sum_{m=-\infty}^{\infty}\frac{\hbar\omega_{c}}{\omega
_{m}^{2}+\varepsilon_{n,\chi}^{2}+\Delta^{2}(T,\nu_{\chi})}\ .
\end{equation}
For $T\rightarrow T_{c}$, $\Delta(T_{c},\nu_{\chi})\rightarrow0$, we can
expand the gap equation with respect to $\Delta^{2}(T_{c},\nu_{\chi})$. To the
first order in $\Delta^{2}(T_{c},\nu_{\chi})$, we arrive at
\begin{align}
1  &  =\frac{x}{2}\hbar\omega_{c}\sum_{n}\frac{\tanh[|\varepsilon_{n,\chi
}|/(2k_{B}T)]}{|\varepsilon_{n,\chi}|}\nonumber\\
&  -x\hbar\omega_{c}k_{B}T\sum_{n}\sum_{m=-\infty}^{\infty}\frac{\Delta
^{2}(T,\nu_{\chi})}{(\omega_{m}^{2}+\varepsilon_{n,\chi}^{2})^{2}}\ .
\label{eq_cgap}%
\end{align}
By using Eq.\ (\ref{self_gap}) and replacing the summation in
Eq.\ (\ref{eq_cgap}) with an integral $\sum_{m}\rightarrow\frac{1}{k_{B}T}%
\int\frac{d\epsilon}{2\pi}$, we finally derive the superconducting gap at half
filling factors to be%
\begin{equation}
\Delta(T,\nu_{\chi}^{\mathrm{H}})=(\hbar\upsilon_{\mathrm{F}}/l_{B}%
)^{3/2}\sqrt{1-T/T_{c}}/[2k_{B}T_{c}\gamma_{n_{\mathrm{F}}}^{(3)}]^{1/2}%
\end{equation}
with $\nu_{\chi}^{\mathrm{H}}=$ $gn_{\mathrm{F}}$, where the critical
temperature is $T_{c}=\Delta(0,\nu_{\chi}^{\mathrm{H}})/(2k_{B})$.
Consequently, in the weak coupling limit, $T_{c}\sim\hbar\upsilon_{\mathrm{F}%
}x/(4k_{B}l_{B})=g|U|eB/(4hk_{B})$, is linearly scaled with the interaction
strength $U$ and the amount of strain $B$, which is similar to the case in
strained graphene~\cite{PhysRevLett.111.046604}. In fact, at partial filling
of the LLs, $T_{c}\propto x$ in the $x\rightarrow0$ limit, while, at the
integer fillings, the transition is quantum critical below the critical
coupling $x_{c}$, as seen in Fig.\ \ref{FigTc}(b). The behavior of the
critical coupling here is different from that for strained graphene in Ref.
\cite{PhysRevLett.111.046604}. For strained graphene, the critical coupling
strength $x_{c}$ distributes symmetrically with respect to the $n=0$ LL, due
to the symmetrically-distributed LLs. For strained \textrm{MoS}$_{2}$
monolayer, however, the $0$-th LL no longer locates at the symmetry point
$\mu=0$, and as a result, the critical coupling is asymmetric for $\nu_{\chi
}=1/2$ and $\nu_{\chi}=-1/2$, as reflected by the critical temperature in Fig.
\ref{FigTc}(b). From Fig. \ref{FigTc}(b), we can also find that the critical
coupling decreases monotonously with the LLs' spacing, which is consistent
with the analytical result presented in Eq.\ (\ref{CriXc}). The linearly
scaled property of the critical temperature with $x$ is distinct from that in
conventional weak coupling superconductors, where $T_{c}\sim\exp(-1/x)$
decreases exponentially with the effective coupling. With the coupling $x$
further increased, the system would cross over to the strong coupling regime
$T_{c}\gtrsim\hbar\omega_{c}$.

\begin{figure}[ptb]
\centering
%Requires \usepackage{graphicx}
\includegraphics[width=\linewidth]{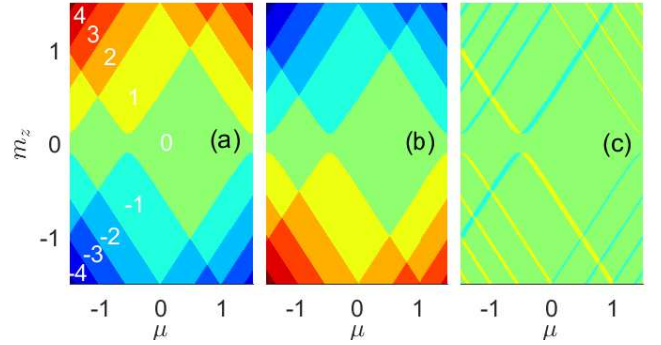} \caption{{}The topological index
(a) $C_{1}^{+\uparrow}$, (b) $C_{1}^{+\downarrow}$ and (c) their summation
$C_{1}^{+} = C_{1}^{+\uparrow}+C_{1}^{+\downarrow}$ as functions of the
chemical potential $\mu$ and Zeeman field $m_{z}$, for $\Delta=0.1\Delta_{0}$.
$C_{1}^{+}$ takes values 0 (green area), +1 (yellow area) and -1 (blue area).
Due to the valley symmetry $E_{n,\overline{\lambda}}^{\overline{\xi}%
\overline{\sigma}}=-E_{n,\lambda}^{\xi\sigma}$, the total topological index
$C_{1}=C_{1}^{+}+C_{1}^{-}$ is of the same shape as (c), but takes values 0
(green area), or $\pm2$ (yellow and blue area).}%
\label{FigCN}%
\end{figure}

\section{Topological Index and Phase Diagram}

\label{TIPD}

In the presence of the Zeeman field, the TR symmetry is broken for the system.
The BdG Hamiltonian without TR symmetry belongs to class
D\cite{PhysRevB.93.054502,PhysRevB.78.195125} and is characterized in 2D by a
topological invariant $C_{1}$, termed as the first Chern number. The
topological index $C_{1}^{\xi\sigma}$ here can be calculated using the Kubo
formula\cite{PhysRevB.90.115305}
\begin{align}
C_{1}^{\xi\sigma}  &  =\frac{i2\pi h^{2}}{A}\sum_{mn}\sum_{k,\lambda
\lambda^{\prime}}(f_{m,\lambda}^{\xi\sigma}-f_{n,\lambda^{\prime}}^{\xi\sigma
})\nonumber\\
&  \times\frac{\langle\Phi_{m,\lambda}^{\xi\sigma}|\upsilon_{x,\xi\sigma}%
|\Phi_{n,\lambda^{\prime}}^{\xi\sigma}\rangle\langle\Phi_{n,\lambda^{\prime}%
}^{\xi\sigma}|\upsilon_{y,\xi\sigma}|\Phi_{m,\lambda}^{\xi\sigma}\rangle
}{(E_{m,\lambda}^{\xi\sigma}-E_{n,\lambda^{\prime}}^{\xi\sigma})^{2}}\ ,
\end{align}
where $\upsilon_{x(y),\xi\sigma}=\hbar^{-1}\partial\widetilde{\mathcal{H}%
}_{\mathrm{BdG}}/\partial k_{x(y)}$, $f_{n,\lambda}^{\xi\sigma}=f(E_{n,\lambda
}^{\xi\sigma})$ represents the Fermi-Dirac distribution function, and%
\begin{equation}
E_{n,\lambda}^{\xi\sigma}=\lambda\sqrt{(\Omega_{\xi\sigma}^{n}-\mu_{\xi\sigma
})^{2}+\Delta^{2}}+\sigma m_{\mathrm{z}}\ , \label{energy}%
\end{equation}%
\begin{equation}
\Phi_{n,\lambda}^{\xi\sigma}=\frac{1}{\sqrt{2}}\left(
\begin{array}
[c]{c}%
\beta_{\xi\sigma,+}^{(n,\lambda)}\\
\lambda\beta_{\xi\sigma,-}^{(n,\lambda)}%
\end{array}
\right)  \otimes\psi_{k,\xi\sigma}^{(n)}\ ,
\end{equation}
are eigenenergy and wavefunction of Hamiltonian (\ref{HBdGb}), respectively,
with $\lambda=\pm$ being the band index and
\begin{equation}
\beta_{\xi\sigma,\pm}^{(n,\lambda)}=\sqrt{1\pm(\Omega_{\xi\sigma}^{n}-\mu
_{\xi\sigma})/(E_{n,\lambda}^{\xi\sigma}-\sigma m_{\mathrm{z}})}\ .
\end{equation}
By using the expressions above, we can derive the topological index to be%
\begin{align}
C_{1}^{\xi\sigma}  &  =\xi\sum_{n=0}^{n_{D}}[\frac{\lambda_{\xi\sigma}}%
{2}(\frac{f_{n+1}^{\xi\sigma}-f_{-n-1}^{\xi\sigma}}{\Omega_{\xi\sigma}^{n+1}%
}-\frac{f_{n}^{\xi\sigma}-f_{-n}^{\xi\sigma}}{\Omega_{\xi\sigma}^{n}%
})\nonumber\\
&  +(n+1/2)(f_{n}^{\xi\sigma}-f_{n+1}^{\xi\sigma}+f_{-n}^{\xi\sigma}%
-f_{-n-1}^{\xi\sigma})]\ , \label{sigxy}%
\end{align}
where $f_{n}^{\xi\sigma}=f_{n,+}^{\xi\sigma}+f_{n,-}^{\xi\sigma}$. The first
line of Eq.\ (\ref{sigxy}) vanishes after the summation over $n$ and, finally,
$C_{1}^{\xi\sigma}$ can be rewritten as
\begin{align}
C_{1}^{\xi\sigma}  &  =\xi\sum_{n=0}^{n_{D}}(n+\frac{1}{2})[(f_{n,+}%
^{\xi\sigma}-f_{n+1,+}^{\xi\sigma}+f_{-n,+}^{\xi\sigma}-f_{-n-1,+}^{\xi\sigma
})\nonumber\\
&  -(f_{-n-1,-}^{\xi\sigma}-f_{-n,-}^{\xi\sigma}+f_{n+1,-}^{\xi\sigma}%
-f_{n,-}^{\xi\sigma})]\ . \label{sigxyf}%
\end{align}
At zero or very low temperatures and for the case of $m_{\mathrm{z}}=0$,
according to Eq.\ (\ref{energy}), $f_{n,+}^{\xi\sigma}=0$ and $f_{n,-}%
^{\xi\sigma}=1$, such that $C_{1}^{\xi\sigma}=0$ always satisfies. However,
for a finite Zeeman field, $m_{\mathrm{z}}\neq0$, by tuning the parameters,
$E_{n,+}^{\xi\sigma}<0$ or $E_{n,-}^{\xi\sigma}>0$ could occur, which changes
$C_{1}^{\xi\sigma}$ from zero to an integer, with the phase boundaries
determined by $m_{\mathrm{z}}^{2}-(\mu_{\xi\sigma}-\Omega_{\xi\sigma}^{n}%
)^{2}=\Delta^{2}$. For example, when $n_{\mathrm{F}}=0$, the topological index
reduces to
\begin{equation}
C_{1}^{\xi\sigma}=(f_{0,+}^{\xi\sigma}+f_{0,-}^{\xi\sigma}-1)\xi\ ,
\end{equation}
such that $C_{1}^{\xi\sigma}=\xi$ ($-\xi$) if $E_{0,+}^{\xi\sigma}<0$
($E_{0,-}^{\xi\sigma}>0$). For $|m_{\mathrm{z}}|>\sqrt{(\mu+\frac{\Delta_{0}%
}{2}+\sigma\xi\lambda_{\mathrm{so}})^{2}+\Delta^{2}}$, $|C_{1}^{\xi\sigma}%
|=1$, which can also be seen from the numerical results displayed in
Figs.\ \ref{FigCN} (a) and (b). As a result, the Chern number for $\xi$ valley
$C_{1}^{\xi}=C_{1}^{\xi\uparrow}+C_{1}^{\xi\downarrow}$ will be $|C_{1}^{\xi
}|=1$, when $(\mu+\frac{\Delta_{0}}{2}+\sigma\xi\lambda_{\mathrm{so}}%
)^{2}<m_{\mathrm{z}}^{2}-\Delta^{2}<(\mu+\frac{\Delta_{0}}{2}-\sigma\xi
\lambda_{\mathrm{so}})^{2}$, as shown in Figs.\ \ref{FigCN} (a) and (b). Odd
values of the Chern number $|C_{1}^{\xi}|$ for the BdG Hamiltonian implies the
emergence of Majorana modes for valley $\xi$. With tuning the chemical
potential, the filling factors will change, and larger values of $C_{1}%
^{\xi\sigma}$ will emerge. However, the difference between $C_{1}^{\xi
\uparrow}$ and $-C_{1}^{\xi\downarrow}$ only can be $0$ or $\pm1$, as shown in
Fig.\ \ref{FigCN}(c). Therefore, $|C_{1}^{\xi}|=0$ and $1$ occur alternately
in the parameter space, as seen from Fig.\ \ref{FigCN} (c). Due to the valley
symmetry $E_{n,\overline{\lambda}}^{\overline{\xi}\overline{\sigma}%
}=-E_{n,\lambda}^{\xi\sigma}$, $C_{1}^{\xi\sigma}=C_{1}^{\overline{\xi
}\overline{\sigma}}$. As a result, the total Chern number for the system
$C_{1}=C_{1}^{+}+C_{1}^{-}=2C_{1}^{\xi}$ is even, as illustrated in
Fig.\ \ref{FigCN} (c). In other words, a pair of Majorana modes emerge
simultaneously in two valleys, when $C_{1}=\pm2$. The realization of odd
values of the total Chern number for the present system requires to further
break the valley symmetry. For example, when the Rashba spin-orbit interaction
is taken into account, the intervalley coupling would destroy the valley
symmetry~\cite{PhysRevB.93.054502}, and then we can expect the emergence of a
single Majorana mode for the edges of the present system.

We have assumed that the $s$-wave pairing for the convenience of calculation.
As proposed in Ref.\ \cite{PhysRevLett.111.046604}, in the presence of
substrates, superconductivity can be triggered by conventional electron-phonon
coupling, meaning that the $s$-wave pairing can possibly be realized in such a
system. Our conclusions, being insensitive to the phase information of the
pairing potential, may not be limited to the s-wave pairing. There could
possibly be other
mechanism\cite{RevModPhys.84.1067,PhysRevLett.98.146801,PhysRevB.75.134512,PhysRevLett.100.146404,PhysRevB.83.220503}%
, such as density wave\cite{PhysRevLett.100.146404}, leading to the
superconductivity, but it may not affect our discussions on the topological
properties of the superconductivity phase.

Theoretically, a topological superconductor in two-dimensions with odd integer
Chern numbers is predicted to host topologically protected gapless chiral
Majorana edge modes. In experiments, while intensive efforts have been made to
search for the chiral Majorana edge modes, the exclusive signature for such
exotic fermions is still under debate. Very recently, following the
theoretical proposal in Ref.\cite{PhysRevB.92.064520}, He $et$ $al.$%
~\cite{He294} have observed the characteristic half-integer longitudinal
conductance. However, the half-integer longitudinal conductance, as argued by
Ji $et$ $al.$~\cite{PhysRevLett.120.107002}, is only a necessary condition for
identifying the Majorana edge modes, but not a sufficient condition. Here, we
propose that the Majorana edge modes can be realized by splitting of the
pseudo LLs. The strained \textrm{MoS}$_{2}$ monolayer can be realized by using
the method proposed in Ref.~\cite{Levy544}. By placing the grown sample on the
setup proposed in Refs.~\cite{PhysRevB.92.064520,He294}, we can expect to
observe an integer to half-integer transition of the longitudinal conductance,
when the Zeeman field is turned on gradually, for an appropriate chemical
potential. The half-integer longitudinal conductance will be a valuable
signature, but possibly not an exclusive evidence, of the Majorana edge modes.
Searching for unambiguous experimental fingerprint of the Majorana edge modes
is still a challenging task at the research front in the condensed-matter physics.

\section{summary}

\label{SUM}

In summary, we have investigated the superconductivity and topological
properties in strained dichalcogenides. We generalized the superconducting
theory for gapless graphene to dichalcogenides with an intrinsic band gap. It
is found that superconductivity can emerge in the pseudo LLs induced by
strain. In the weak coupling limit, the superconducting gap is linearly-scaled
with the coupling strength for the partial fillings, in contrast to
conventional weak coupling superconductors. The superconductivity gap is
maximized when the LLs are half-filled, but for integer fillings the
superconductivity is fully suppressed, when the coupling strength is below a
quantum critical value. We find the quantum critical coupling strength is
determined by the spacing between the two LLs closest to the Fermi level.
Interestingly, in the presence of a Zeeman field, a pair of Majorana modes
emerge simultaneously in the two valleys of strained dichalcogenides. A single
Majorana mode can be realized, if the valley symmetry is further lifted.

\section{acknowledgements}

We thank Prof. Tao Zhou for helpful discussions. This work was supported by
the State Key Program for Basic Researches of China under Grants No.
2015CB921202, and No. 2017YFA0303203 (D.Y.X), the National Natural Science
Foundation of China under Grants No. 11674160 (L.S.), No. 11574155 (R.M.), No.
11474149 (R.S.), No. 11804130 (W.L.), No. 11474106 (R.-Q.W) and the Key
Program for Guangdong NSF of China under Grant No. 2017B030311003 (R.-Q.W) and
GDUPS (2017).

\bibliography{bibref}

\begin{thebibliography}{44}
\expandafter\ifx\csname natexlab\endcsname\relax\def\natexlab#1{#1}\fi
\expandafter\ifx\csname bibnamefont\endcsname\relax
  \def\bibnamefont#1{#1}\fi
\expandafter\ifx\csname bibfnamefont\endcsname\relax
  \def\bibfnamefont#1{#1}\fi
\expandafter\ifx\csname citenamefont\endcsname\relax
  \def\citenamefont#1{#1}\fi
\expandafter\ifx\csname url\endcsname\relax
  \def\url#1{\texttt{#1}}\fi
\expandafter\ifx\csname urlprefix\endcsname\relax\def\urlprefix{URL }\fi
\providecommand{\bibinfo}[2]{#2}
\providecommand{\eprint}[2][]{\url{#2}}

\bibitem[{\citenamefont{Novoselov et~al.}(2004)\citenamefont{Novoselov, Geim,
  Morozov, Jiang, Zhang, Dubonos, Grigorieva, and Firsov}}]{Novoselov666}
\bibinfo{author}{\bibfnamefont{K.~S.} \bibnamefont{Novoselov}},
  \bibinfo{author}{\bibfnamefont{A.~K.} \bibnamefont{Geim}},
  \bibinfo{author}{\bibfnamefont{S.~V.} \bibnamefont{Morozov}},
  \bibinfo{author}{\bibfnamefont{D.}~\bibnamefont{Jiang}},
  \bibinfo{author}{\bibfnamefont{Y.}~\bibnamefont{Zhang}},
  \bibinfo{author}{\bibfnamefont{S.~V.} \bibnamefont{Dubonos}},
  \bibinfo{author}{\bibfnamefont{I.~V.} \bibnamefont{Grigorieva}},
  \bibnamefont{and} \bibinfo{author}{\bibfnamefont{A.~A.}
  \bibnamefont{Firsov}}, \bibinfo{journal}{Science}
  \textbf{\bibinfo{volume}{306}}, \bibinfo{pages}{666} (\bibinfo{year}{2004}).

\bibitem[{\citenamefont{Novoselov
  et~al.}(2005{\natexlab{a}})\citenamefont{Novoselov, Geim, Morozov, Jiang,
  Katsnelson, Grigorieva, Dubonos, and Firsov}}]{Novoselov:2005aa}
\bibinfo{author}{\bibfnamefont{K.~S.} \bibnamefont{Novoselov}},
  \bibinfo{author}{\bibfnamefont{A.~K.} \bibnamefont{Geim}},
  \bibinfo{author}{\bibfnamefont{S.~V.} \bibnamefont{Morozov}},
  \bibinfo{author}{\bibfnamefont{D.}~\bibnamefont{Jiang}},
  \bibinfo{author}{\bibfnamefont{M.~I.} \bibnamefont{Katsnelson}},
  \bibinfo{author}{\bibfnamefont{I.~V.} \bibnamefont{Grigorieva}},
  \bibinfo{author}{\bibfnamefont{S.~V.} \bibnamefont{Dubonos}},
  \bibnamefont{and} \bibinfo{author}{\bibfnamefont{A.~A.}
  \bibnamefont{Firsov}}, \bibinfo{journal}{Nature}
  \textbf{\bibinfo{volume}{438}}, \bibinfo{pages}{197}
  (\bibinfo{year}{2005}{\natexlab{a}}).

\bibitem[{\citenamefont{Zhang et~al.}(2005)\citenamefont{Zhang, Tan, Stormer,
  and Kim}}]{Zhang:2005aa}
\bibinfo{author}{\bibfnamefont{Y.}~\bibnamefont{Zhang}},
  \bibinfo{author}{\bibfnamefont{Y.-W.} \bibnamefont{Tan}},
  \bibinfo{author}{\bibfnamefont{H.~L.} \bibnamefont{Stormer}},
  \bibnamefont{and} \bibinfo{author}{\bibfnamefont{P.}~\bibnamefont{Kim}},
  \bibinfo{journal}{Nature} \textbf{\bibinfo{volume}{438}},
  \bibinfo{pages}{201} (\bibinfo{year}{2005}).

\bibitem[{\citenamefont{Novoselov
  et~al.}(2005{\natexlab{b}})\citenamefont{Novoselov, Jiang, Schedin, Booth,
  Khotkevich, Morozov, and Geim}}]{Novoselov10451}
\bibinfo{author}{\bibfnamefont{K.~S.} \bibnamefont{Novoselov}},
  \bibinfo{author}{\bibfnamefont{D.}~\bibnamefont{Jiang}},
  \bibinfo{author}{\bibfnamefont{F.}~\bibnamefont{Schedin}},
  \bibinfo{author}{\bibfnamefont{T.~J.} \bibnamefont{Booth}},
  \bibinfo{author}{\bibfnamefont{V.~V.} \bibnamefont{Khotkevich}},
  \bibinfo{author}{\bibfnamefont{S.~V.} \bibnamefont{Morozov}},
  \bibnamefont{and} \bibinfo{author}{\bibfnamefont{A.~K.} \bibnamefont{Geim}},
  \bibinfo{journal}{Proceedings of the National Academy of Sciences}
  \textbf{\bibinfo{volume}{102}}, \bibinfo{pages}{10451}
  (\bibinfo{year}{2005}{\natexlab{b}}).

\bibitem[{\citenamefont{Lee et~al.}(2010)\citenamefont{Lee, Li, Kalb, Liu,
  Berger, Carpick, and Hone}}]{Lee76}
\bibinfo{author}{\bibfnamefont{C.}~\bibnamefont{Lee}},
  \bibinfo{author}{\bibfnamefont{Q.}~\bibnamefont{Li}},
  \bibinfo{author}{\bibfnamefont{W.}~\bibnamefont{Kalb}},
  \bibinfo{author}{\bibfnamefont{X.-Z.} \bibnamefont{Liu}},
  \bibinfo{author}{\bibfnamefont{H.}~\bibnamefont{Berger}},
  \bibinfo{author}{\bibfnamefont{R.~W.} \bibnamefont{Carpick}},
  \bibnamefont{and} \bibinfo{author}{\bibfnamefont{J.}~\bibnamefont{Hone}},
  \bibinfo{journal}{Science} \textbf{\bibinfo{volume}{328}},
  \bibinfo{pages}{76} (\bibinfo{year}{2010}).

\bibitem[{\citenamefont{Goerbig}(2011)}]{RevModPhys.83.1193}
\bibinfo{author}{\bibfnamefont{M.~O.} \bibnamefont{Goerbig}},
  \bibinfo{journal}{Rev. Mod. Phys.} \textbf{\bibinfo{volume}{83}},
  \bibinfo{pages}{1193} (\bibinfo{year}{2011}).

\bibitem[{\citenamefont{Kotov et~al.}(2012)\citenamefont{Kotov, Uchoa, Pereira,
  Guinea, and Castro~Neto}}]{RevModPhys.84.1067}
\bibinfo{author}{\bibfnamefont{V.~N.} \bibnamefont{Kotov}},
  \bibinfo{author}{\bibfnamefont{B.}~\bibnamefont{Uchoa}},
  \bibinfo{author}{\bibfnamefont{V.~M.} \bibnamefont{Pereira}},
  \bibinfo{author}{\bibfnamefont{F.}~\bibnamefont{Guinea}}, \bibnamefont{and}
  \bibinfo{author}{\bibfnamefont{A.~H.} \bibnamefont{Castro~Neto}},
  \bibinfo{journal}{Rev. Mod. Phys.} \textbf{\bibinfo{volume}{84}},
  \bibinfo{pages}{1067} (\bibinfo{year}{2012}).

\bibitem[{\citenamefont{Gomes et~al.}(2012)\citenamefont{Gomes, Mar, Ko,
  Guinea, and Manoharan}}]{Gomes:2012aa}
\bibinfo{author}{\bibfnamefont{K.~K.} \bibnamefont{Gomes}},
  \bibinfo{author}{\bibfnamefont{W.}~\bibnamefont{Mar}},
  \bibinfo{author}{\bibfnamefont{W.}~\bibnamefont{Ko}},
  \bibinfo{author}{\bibfnamefont{F.}~\bibnamefont{Guinea}}, \bibnamefont{and}
  \bibinfo{author}{\bibfnamefont{H.~C.} \bibnamefont{Manoharan}},
  \bibinfo{journal}{Nature} \textbf{\bibinfo{volume}{483}},
  \bibinfo{pages}{306} (\bibinfo{year}{2012}).

\bibitem[{\citenamefont{Gunawan et~al.}(2006)\citenamefont{Gunawan, Shkolnikov,
  Vakili, Gokmen, De~Poortere, and Shayegan}}]{PhysRevLett.97.186404}
\bibinfo{author}{\bibfnamefont{O.}~\bibnamefont{Gunawan}},
  \bibinfo{author}{\bibfnamefont{Y.~P.} \bibnamefont{Shkolnikov}},
  \bibinfo{author}{\bibfnamefont{K.}~\bibnamefont{Vakili}},
  \bibinfo{author}{\bibfnamefont{T.}~\bibnamefont{Gokmen}},
  \bibinfo{author}{\bibfnamefont{E.~P.} \bibnamefont{De~Poortere}},
  \bibnamefont{and} \bibinfo{author}{\bibfnamefont{M.}~\bibnamefont{Shayegan}},
  \bibinfo{journal}{Phys. Rev. Lett.} \textbf{\bibinfo{volume}{97}},
  \bibinfo{pages}{186404} (\bibinfo{year}{2006}).

\bibitem[{\citenamefont{Yao et~al.}(2008)\citenamefont{Yao, Xiao, and
  Niu}}]{PhysRevB.77.235406}
\bibinfo{author}{\bibfnamefont{W.}~\bibnamefont{Yao}},
  \bibinfo{author}{\bibfnamefont{D.}~\bibnamefont{Xiao}}, \bibnamefont{and}
  \bibinfo{author}{\bibfnamefont{Q.}~\bibnamefont{Niu}},
  \bibinfo{journal}{Phys. Rev. B} \textbf{\bibinfo{volume}{77}},
  \bibinfo{pages}{235406} (\bibinfo{year}{2008}).

\bibitem[{\citenamefont{Xiao et~al.}(2007)\citenamefont{Xiao, Yao, and
  Niu}}]{PhysRevLett.99.236809}
\bibinfo{author}{\bibfnamefont{D.}~\bibnamefont{Xiao}},
  \bibinfo{author}{\bibfnamefont{W.}~\bibnamefont{Yao}}, \bibnamefont{and}
  \bibinfo{author}{\bibfnamefont{Q.}~\bibnamefont{Niu}},
  \bibinfo{journal}{Phys. Rev. Lett.} \textbf{\bibinfo{volume}{99}},
  \bibinfo{pages}{236809} (\bibinfo{year}{2007}).

\bibitem[{\citenamefont{Rycerz et~al.}(2007)\citenamefont{Rycerz, Tworzyd{\l}o,
  and Beenakker}}]{Rycerz:2007aa}
\bibinfo{author}{\bibfnamefont{A.}~\bibnamefont{Rycerz}},
  \bibinfo{author}{\bibfnamefont{J.}~\bibnamefont{Tworzyd{\l}o}},
  \bibnamefont{and} \bibinfo{author}{\bibfnamefont{C.~W.~J.}
  \bibnamefont{Beenakker}}, \bibinfo{journal}{Nature Physics}
  \textbf{\bibinfo{volume}{3}}, \bibinfo{pages}{172} (\bibinfo{year}{2007}).

\bibitem[{\citenamefont{Zhang et~al.}(2011)\citenamefont{Zhang, Jung, Fiete,
  Niu, and MacDonald}}]{PhysRevLett.106.156801}
\bibinfo{author}{\bibfnamefont{F.}~\bibnamefont{Zhang}},
  \bibinfo{author}{\bibfnamefont{J.}~\bibnamefont{Jung}},
  \bibinfo{author}{\bibfnamefont{G.~A.} \bibnamefont{Fiete}},
  \bibinfo{author}{\bibfnamefont{Q.}~\bibnamefont{Niu}}, \bibnamefont{and}
  \bibinfo{author}{\bibfnamefont{A.~H.} \bibnamefont{MacDonald}},
  \bibinfo{journal}{Phys. Rev. Lett.} \textbf{\bibinfo{volume}{106}},
  \bibinfo{pages}{156801} (\bibinfo{year}{2011}).

\bibitem[{\citenamefont{Min et~al.}(2006)\citenamefont{Min, Hill, Sinitsyn,
  Sahu, Kleinman, and MacDonald}}]{PhysRevB.74.165310}
\bibinfo{author}{\bibfnamefont{H.}~\bibnamefont{Min}},
  \bibinfo{author}{\bibfnamefont{J.~E.} \bibnamefont{Hill}},
  \bibinfo{author}{\bibfnamefont{N.~A.} \bibnamefont{Sinitsyn}},
  \bibinfo{author}{\bibfnamefont{B.~R.} \bibnamefont{Sahu}},
  \bibinfo{author}{\bibfnamefont{L.}~\bibnamefont{Kleinman}}, \bibnamefont{and}
  \bibinfo{author}{\bibfnamefont{A.~H.} \bibnamefont{MacDonald}},
  \bibinfo{journal}{Phys. Rev. B} \textbf{\bibinfo{volume}{74}},
  \bibinfo{pages}{165310} (\bibinfo{year}{2006}).

\bibitem[{\citenamefont{Yao et~al.}(2007)\citenamefont{Yao, Ye, Qi, Zhang, and
  Fang}}]{PhysRevB.75.041401}
\bibinfo{author}{\bibfnamefont{Y.}~\bibnamefont{Yao}},
  \bibinfo{author}{\bibfnamefont{F.}~\bibnamefont{Ye}},
  \bibinfo{author}{\bibfnamefont{X.-L.} \bibnamefont{Qi}},
  \bibinfo{author}{\bibfnamefont{S.-C.} \bibnamefont{Zhang}}, \bibnamefont{and}
  \bibinfo{author}{\bibfnamefont{Z.}~\bibnamefont{Fang}},
  \bibinfo{journal}{Phys. Rev. B} \textbf{\bibinfo{volume}{75}},
  \bibinfo{pages}{041401} (\bibinfo{year}{2007}).

\bibitem[{\citenamefont{Avsar et~al.}(2014)\citenamefont{Avsar, Tan,
  Taychatanapat, Balakrishnan, Koon, Yeo, Lahiri, Carvalho, Rodin, O'Farrell
  et~al.}}]{Avsar:2014aa}
\bibinfo{author}{\bibfnamefont{A.}~\bibnamefont{Avsar}},
  \bibinfo{author}{\bibfnamefont{J.~Y.} \bibnamefont{Tan}},
  \bibinfo{author}{\bibfnamefont{T.}~\bibnamefont{Taychatanapat}},
  \bibinfo{author}{\bibfnamefont{J.}~\bibnamefont{Balakrishnan}},
  \bibinfo{author}{\bibfnamefont{G.~K.~W.} \bibnamefont{Koon}},
  \bibinfo{author}{\bibfnamefont{Y.}~\bibnamefont{Yeo}},
  \bibinfo{author}{\bibfnamefont{J.}~\bibnamefont{Lahiri}},
  \bibinfo{author}{\bibfnamefont{A.}~\bibnamefont{Carvalho}},
  \bibinfo{author}{\bibfnamefont{A.~S.} \bibnamefont{Rodin}},
  \bibinfo{author}{\bibfnamefont{E.~C.~T.} \bibnamefont{O'Farrell}},
  \bibnamefont{et~al.}, \bibinfo{journal}{Nature Communications}
  \textbf{\bibinfo{volume}{5}}, \bibinfo{pages}{4875} (\bibinfo{year}{2014}).

\bibitem[{\citenamefont{Cummings et~al.}(2017)\citenamefont{Cummings, Garcia,
  Fabian, and Roche}}]{PhysRevLett.119.206601}
\bibinfo{author}{\bibfnamefont{A.~W.} \bibnamefont{Cummings}},
  \bibinfo{author}{\bibfnamefont{J.~H.} \bibnamefont{Garcia}},
  \bibinfo{author}{\bibfnamefont{J.}~\bibnamefont{Fabian}}, \bibnamefont{and}
  \bibinfo{author}{\bibfnamefont{S.}~\bibnamefont{Roche}},
  \bibinfo{journal}{Phys. Rev. Lett.} \textbf{\bibinfo{volume}{119}},
  \bibinfo{pages}{206601} (\bibinfo{year}{2017}).

\bibitem[{\citenamefont{Levy et~al.}(2010)\citenamefont{Levy, Burke, Meaker,
  Panlasigui, Zettl, Guinea, Neto, and Crommie}}]{Levy544}
\bibinfo{author}{\bibfnamefont{N.}~\bibnamefont{Levy}},
  \bibinfo{author}{\bibfnamefont{S.~A.} \bibnamefont{Burke}},
  \bibinfo{author}{\bibfnamefont{K.~L.} \bibnamefont{Meaker}},
  \bibinfo{author}{\bibfnamefont{M.}~\bibnamefont{Panlasigui}},
  \bibinfo{author}{\bibfnamefont{A.}~\bibnamefont{Zettl}},
  \bibinfo{author}{\bibfnamefont{F.}~\bibnamefont{Guinea}},
  \bibinfo{author}{\bibfnamefont{A.~H.~C.} \bibnamefont{Neto}},
  \bibnamefont{and} \bibinfo{author}{\bibfnamefont{M.~F.}
  \bibnamefont{Crommie}}, \bibinfo{journal}{Science}
  \textbf{\bibinfo{volume}{329}}, \bibinfo{pages}{544} (\bibinfo{year}{2010}).

\bibitem[{\citenamefont{Abanin and Pesin}(2012)}]{PhysRevLett.109.066802}
\bibinfo{author}{\bibfnamefont{D.~A.} \bibnamefont{Abanin}} \bibnamefont{and}
  \bibinfo{author}{\bibfnamefont{D.~A.} \bibnamefont{Pesin}},
  \bibinfo{journal}{Phys. Rev. Lett.} \textbf{\bibinfo{volume}{109}},
  \bibinfo{pages}{066802} (\bibinfo{year}{2012}).

\bibitem[{\citenamefont{Ghaemi et~al.}(2012)\citenamefont{Ghaemi, Cayssol,
  Sheng, and Vishwanath}}]{PhysRevLett.108.266801}
\bibinfo{author}{\bibfnamefont{P.}~\bibnamefont{Ghaemi}},
  \bibinfo{author}{\bibfnamefont{J.}~\bibnamefont{Cayssol}},
  \bibinfo{author}{\bibfnamefont{D.~N.} \bibnamefont{Sheng}}, \bibnamefont{and}
  \bibinfo{author}{\bibfnamefont{A.}~\bibnamefont{Vishwanath}},
  \bibinfo{journal}{Phys. Rev. Lett.} \textbf{\bibinfo{volume}{108}},
  \bibinfo{pages}{266801} (\bibinfo{year}{2012}).

\bibitem[{\citenamefont{Uchoa and Barlas}(2013)}]{PhysRevLett.111.046604}
\bibinfo{author}{\bibfnamefont{B.}~\bibnamefont{Uchoa}} \bibnamefont{and}
  \bibinfo{author}{\bibfnamefont{Y.}~\bibnamefont{Barlas}},
  \bibinfo{journal}{Phys. Rev. Lett.} \textbf{\bibinfo{volume}{111}},
  \bibinfo{pages}{046604} (\bibinfo{year}{2013}).

\bibitem[{\citenamefont{Wang and Wu}(2016)}]{PhysRevB.93.054502}
\bibinfo{author}{\bibfnamefont{L.}~\bibnamefont{Wang}} \bibnamefont{and}
  \bibinfo{author}{\bibfnamefont{M.~W.} \bibnamefont{Wu}},
  \bibinfo{journal}{Phys. Rev. B} \textbf{\bibinfo{volume}{93}},
  \bibinfo{pages}{054502} (\bibinfo{year}{2016}).

\bibitem[{\citenamefont{Splendiani et~al.}(2010)\citenamefont{Splendiani, Sun,
  Zhang, Li, Kim, Chim, Galli, and Wang}}]{nl903868w}
\bibinfo{author}{\bibfnamefont{A.}~\bibnamefont{Splendiani}},
  \bibinfo{author}{\bibfnamefont{L.}~\bibnamefont{Sun}},
  \bibinfo{author}{\bibfnamefont{Y.}~\bibnamefont{Zhang}},
  \bibinfo{author}{\bibfnamefont{T.}~\bibnamefont{Li}},
  \bibinfo{author}{\bibfnamefont{J.}~\bibnamefont{Kim}},
  \bibinfo{author}{\bibfnamefont{C.-Y.} \bibnamefont{Chim}},
  \bibinfo{author}{\bibfnamefont{G.}~\bibnamefont{Galli}}, \bibnamefont{and}
  \bibinfo{author}{\bibfnamefont{F.}~\bibnamefont{Wang}},
  \bibinfo{journal}{Nano Letters} \textbf{\bibinfo{volume}{10}},
  \bibinfo{pages}{1271} (\bibinfo{year}{2010}).

\bibitem[{\citenamefont{Mak et~al.}(2010)\citenamefont{Mak, Lee, Hone, Shan,
  and Heinz}}]{PhysRevLett.105.136805}
\bibinfo{author}{\bibfnamefont{K.~F.} \bibnamefont{Mak}},
  \bibinfo{author}{\bibfnamefont{C.}~\bibnamefont{Lee}},
  \bibinfo{author}{\bibfnamefont{J.}~\bibnamefont{Hone}},
  \bibinfo{author}{\bibfnamefont{J.}~\bibnamefont{Shan}}, \bibnamefont{and}
  \bibinfo{author}{\bibfnamefont{T.~F.} \bibnamefont{Heinz}},
  \bibinfo{journal}{Phys. Rev. Lett.} \textbf{\bibinfo{volume}{105}},
  \bibinfo{pages}{136805} (\bibinfo{year}{2010}).

\bibitem[{\citenamefont{Radisavljevic et~al.}(2011)\citenamefont{Radisavljevic,
  Radenovic, Brivio, Giacometti, and Kis}}]{Radisavljevic:2011aa}
\bibinfo{author}{\bibfnamefont{B.}~\bibnamefont{Radisavljevic}},
  \bibinfo{author}{\bibfnamefont{A.}~\bibnamefont{Radenovic}},
  \bibinfo{author}{\bibfnamefont{J.}~\bibnamefont{Brivio}},
  \bibinfo{author}{\bibfnamefont{V.}~\bibnamefont{Giacometti}},
  \bibnamefont{and} \bibinfo{author}{\bibfnamefont{A.}~\bibnamefont{Kis}},
  \bibinfo{journal}{Nature Nanotechnology} \textbf{\bibinfo{volume}{6}},
  \bibinfo{pages}{147} (\bibinfo{year}{2011}).

\bibitem[{\citenamefont{Korn et~al.}(2011)\citenamefont{Korn, Heydrich, Hirmer,
  Schmutzler, and Sch{\"u}ller}}]{aip_apl99}
\bibinfo{author}{\bibfnamefont{T.}~\bibnamefont{Korn}},
  \bibinfo{author}{\bibfnamefont{S.}~\bibnamefont{Heydrich}},
  \bibinfo{author}{\bibfnamefont{M.}~\bibnamefont{Hirmer}},
  \bibinfo{author}{\bibfnamefont{J.}~\bibnamefont{Schmutzler}},
  \bibnamefont{and}
  \bibinfo{author}{\bibfnamefont{C.}~\bibnamefont{Sch{\"u}ller}},
  \bibinfo{journal}{Applied Physics Letters} \textbf{\bibinfo{volume}{99}},
  \bibinfo{pages}{102109} (\bibinfo{year}{2011}).

\bibitem[{\citenamefont{Wakamura et~al.}(2018)\citenamefont{Wakamura, Reale,
  Palczynski, Gu\'eron, Mattevi, and Bouchiat}}]{PhysRevLett.120.106802}
\bibinfo{author}{\bibfnamefont{T.}~\bibnamefont{Wakamura}},
  \bibinfo{author}{\bibfnamefont{F.}~\bibnamefont{Reale}},
  \bibinfo{author}{\bibfnamefont{P.}~\bibnamefont{Palczynski}},
  \bibinfo{author}{\bibfnamefont{S.}~\bibnamefont{Gu\'eron}},
  \bibinfo{author}{\bibfnamefont{C.}~\bibnamefont{Mattevi}}, \bibnamefont{and}
  \bibinfo{author}{\bibfnamefont{H.}~\bibnamefont{Bouchiat}},
  \bibinfo{journal}{Phys. Rev. Lett.} \textbf{\bibinfo{volume}{120}},
  \bibinfo{pages}{106802} (\bibinfo{year}{2018}).

\bibitem[{\citenamefont{Zihlmann et~al.}(2018)\citenamefont{Zihlmann, Cummings,
  Garcia, Kedves, Watanabe, Taniguchi, Sch\"onenberger, and
  Makk}}]{PhysRevB.97.075434}
\bibinfo{author}{\bibfnamefont{S.}~\bibnamefont{Zihlmann}},
  \bibinfo{author}{\bibfnamefont{A.~W.} \bibnamefont{Cummings}},
  \bibinfo{author}{\bibfnamefont{J.~H.} \bibnamefont{Garcia}},
  \bibinfo{author}{\bibfnamefont{M.}~\bibnamefont{Kedves}},
  \bibinfo{author}{\bibfnamefont{K.}~\bibnamefont{Watanabe}},
  \bibinfo{author}{\bibfnamefont{T.}~\bibnamefont{Taniguchi}},
  \bibinfo{author}{\bibfnamefont{C.}~\bibnamefont{Sch\"onenberger}},
  \bibnamefont{and} \bibinfo{author}{\bibfnamefont{P.}~\bibnamefont{Makk}},
  \bibinfo{journal}{Phys. Rev. B} \textbf{\bibinfo{volume}{97}},
  \bibinfo{pages}{075434} (\bibinfo{year}{2018}).

\bibitem[{\citenamefont{Wang et~al.}(2012)\citenamefont{Wang, Kalantar-Zadeh,
  Kis, Coleman, and Strano}}]{Wang:2012aa}
\bibinfo{author}{\bibfnamefont{Q.~H.} \bibnamefont{Wang}},
  \bibinfo{author}{\bibfnamefont{K.}~\bibnamefont{Kalantar-Zadeh}},
  \bibinfo{author}{\bibfnamefont{A.}~\bibnamefont{Kis}},
  \bibinfo{author}{\bibfnamefont{J.~N.} \bibnamefont{Coleman}},
  \bibnamefont{and} \bibinfo{author}{\bibfnamefont{M.~S.}
  \bibnamefont{Strano}}, \bibinfo{journal}{Nature Nanotechnology}
  \textbf{\bibinfo{volume}{7}}, \bibinfo{pages}{699} (\bibinfo{year}{2012}).

\bibitem[{\citenamefont{Xiao et~al.}(2012)\citenamefont{Xiao, Liu, Feng, Xu,
  and Yao}}]{PhysRevLett.108.196802}
\bibinfo{author}{\bibfnamefont{D.}~\bibnamefont{Xiao}},
  \bibinfo{author}{\bibfnamefont{G.-B.} \bibnamefont{Liu}},
  \bibinfo{author}{\bibfnamefont{W.}~\bibnamefont{Feng}},
  \bibinfo{author}{\bibfnamefont{X.}~\bibnamefont{Xu}}, \bibnamefont{and}
  \bibinfo{author}{\bibfnamefont{W.}~\bibnamefont{Yao}},
  \bibinfo{journal}{Phys. Rev. Lett.} \textbf{\bibinfo{volume}{108}},
  \bibinfo{pages}{196802} (\bibinfo{year}{2012}).

\bibitem[{\citenamefont{Garcia et~al.}(2017)\citenamefont{Garcia, Cummings, and
  Roche}}]{acs.nanolett.7b02364}
\bibinfo{author}{\bibfnamefont{J.~H.} \bibnamefont{Garcia}},
  \bibinfo{author}{\bibfnamefont{A.~W.} \bibnamefont{Cummings}},
  \bibnamefont{and} \bibinfo{author}{\bibfnamefont{S.}~\bibnamefont{Roche}},
  \bibinfo{journal}{Nano Letters} \textbf{\bibinfo{volume}{17}},
  \bibinfo{pages}{5078} (\bibinfo{year}{2017}).

\bibitem[{\citenamefont{Schmidt et~al.}(2016)\citenamefont{Schmidt, Yudhistira,
  Chu, Castro~Neto, \"Ozyilmaz, Adam, and Eda}}]{PhysRevLett.116.046803}
\bibinfo{author}{\bibfnamefont{H.}~\bibnamefont{Schmidt}},
  \bibinfo{author}{\bibfnamefont{I.}~\bibnamefont{Yudhistira}},
  \bibinfo{author}{\bibfnamefont{L.}~\bibnamefont{Chu}},
  \bibinfo{author}{\bibfnamefont{A.~H.} \bibnamefont{Castro~Neto}},
  \bibinfo{author}{\bibfnamefont{B.}~\bibnamefont{\"Ozyilmaz}},
  \bibinfo{author}{\bibfnamefont{S.}~\bibnamefont{Adam}}, \bibnamefont{and}
  \bibinfo{author}{\bibfnamefont{G.}~\bibnamefont{Eda}},
  \bibinfo{journal}{Phys. Rev. Lett.} \textbf{\bibinfo{volume}{116}},
  \bibinfo{pages}{046803} (\bibinfo{year}{2016}).

\bibitem[{\citenamefont{Zhang et~al.}(2017)\citenamefont{Zhang, Shi, Ye,
  Suzuki, and Iwasa}}]{PhysRevB.95.205302}
\bibinfo{author}{\bibfnamefont{Y.~J.} \bibnamefont{Zhang}},
  \bibinfo{author}{\bibfnamefont{W.}~\bibnamefont{Shi}},
  \bibinfo{author}{\bibfnamefont{J.~T.} \bibnamefont{Ye}},
  \bibinfo{author}{\bibfnamefont{R.}~\bibnamefont{Suzuki}}, \bibnamefont{and}
  \bibinfo{author}{\bibfnamefont{Y.}~\bibnamefont{Iwasa}},
  \bibinfo{journal}{Phys. Rev. B} \textbf{\bibinfo{volume}{95}},
  \bibinfo{pages}{205302} (\bibinfo{year}{2017}).

\bibitem[{\citenamefont{Zhu et~al.}(2011)\citenamefont{Zhu, Cheng, and
  Schwingenschl\"ogl}}]{PhysRevB.84.153402}
\bibinfo{author}{\bibfnamefont{Z.~Y.} \bibnamefont{Zhu}},
  \bibinfo{author}{\bibfnamefont{Y.~C.} \bibnamefont{Cheng}}, \bibnamefont{and}
  \bibinfo{author}{\bibfnamefont{U.}~\bibnamefont{Schwingenschl\"ogl}},
  \bibinfo{journal}{Phys. Rev. B} \textbf{\bibinfo{volume}{84}},
  \bibinfo{pages}{153402} (\bibinfo{year}{2011}).

\bibitem[{\citenamefont{Deng et~al.}(2016)\citenamefont{Deng, Wang, Luo, Sheng,
  Wang, and Xing}}]{1367-2630-18-9-093040}
\bibinfo{author}{\bibfnamefont{M.-X.} \bibnamefont{Deng}},
  \bibinfo{author}{\bibfnamefont{R.-Q.} \bibnamefont{Wang}},
  \bibinfo{author}{\bibfnamefont{W.}~\bibnamefont{Luo}},
  \bibinfo{author}{\bibfnamefont{L.}~\bibnamefont{Sheng}},
  \bibinfo{author}{\bibfnamefont{B.~G.} \bibnamefont{Wang}}, \bibnamefont{and}
  \bibinfo{author}{\bibfnamefont{D.~Y.} \bibnamefont{Xing}},
  \bibinfo{journal}{New Journal of Physics} \textbf{\bibinfo{volume}{18}},
  \bibinfo{pages}{093040} (\bibinfo{year}{2016}).

\bibitem[{\citenamefont{Schnyder et~al.}(2008)\citenamefont{Schnyder, Ryu,
  Furusaki, and Ludwig}}]{PhysRevB.78.195125}
\bibinfo{author}{\bibfnamefont{A.~P.} \bibnamefont{Schnyder}},
  \bibinfo{author}{\bibfnamefont{S.}~\bibnamefont{Ryu}},
  \bibinfo{author}{\bibfnamefont{A.}~\bibnamefont{Furusaki}}, \bibnamefont{and}
  \bibinfo{author}{\bibfnamefont{A.~W.~W.} \bibnamefont{Ludwig}},
  \bibinfo{journal}{Phys. Rev. B} \textbf{\bibinfo{volume}{78}},
  \bibinfo{pages}{195125} (\bibinfo{year}{2008}).

\bibitem[{\citenamefont{Zhang et~al.}(2014)\citenamefont{Zhang, Zhang, and
  Shen}}]{PhysRevB.90.115305}
\bibinfo{author}{\bibfnamefont{S.-B.} \bibnamefont{Zhang}},
  \bibinfo{author}{\bibfnamefont{Y.-Y.} \bibnamefont{Zhang}}, \bibnamefont{and}
  \bibinfo{author}{\bibfnamefont{S.-Q.} \bibnamefont{Shen}},
  \bibinfo{journal}{Phys. Rev. B} \textbf{\bibinfo{volume}{90}},
  \bibinfo{pages}{115305} (\bibinfo{year}{2014}).

\bibitem[{\citenamefont{Uchoa and Castro~Neto}(2007)}]{PhysRevLett.98.146801}
\bibinfo{author}{\bibfnamefont{B.}~\bibnamefont{Uchoa}} \bibnamefont{and}
  \bibinfo{author}{\bibfnamefont{A.~H.} \bibnamefont{Castro~Neto}},
  \bibinfo{journal}{Phys. Rev. Lett.} \textbf{\bibinfo{volume}{98}},
  \bibinfo{pages}{146801} (\bibinfo{year}{2007}).

\bibitem[{\citenamefont{Black-Schaffer and Doniach}(2007)}]{PhysRevB.75.134512}
\bibinfo{author}{\bibfnamefont{A.~M.} \bibnamefont{Black-Schaffer}}
  \bibnamefont{and} \bibinfo{author}{\bibfnamefont{S.}~\bibnamefont{Doniach}},
  \bibinfo{journal}{Phys. Rev. B} \textbf{\bibinfo{volume}{75}},
  \bibinfo{pages}{134512} (\bibinfo{year}{2007}).

\bibitem[{\citenamefont{Honerkamp}(2008)}]{PhysRevLett.100.146404}
\bibinfo{author}{\bibfnamefont{C.}~\bibnamefont{Honerkamp}},
  \bibinfo{journal}{Phys. Rev. Lett.} \textbf{\bibinfo{volume}{100}},
  \bibinfo{pages}{146404} (\bibinfo{year}{2008}).

\bibitem[{\citenamefont{Kopnin et~al.}(2011)\citenamefont{Kopnin, Heikkil\"a,
  and Volovik}}]{PhysRevB.83.220503}
\bibinfo{author}{\bibfnamefont{N.~B.} \bibnamefont{Kopnin}},
  \bibinfo{author}{\bibfnamefont{T.~T.} \bibnamefont{Heikkil\"a}},
  \bibnamefont{and} \bibinfo{author}{\bibfnamefont{G.~E.}
  \bibnamefont{Volovik}}, \bibinfo{journal}{Phys. Rev. B}
  \textbf{\bibinfo{volume}{83}}, \bibinfo{pages}{220503}
  (\bibinfo{year}{2011}).

\bibitem[{\citenamefont{Wang et~al.}(2015)\citenamefont{Wang, Zhou, Lian, and
  Zhang}}]{PhysRevB.92.064520}
\bibinfo{author}{\bibfnamefont{J.}~\bibnamefont{Wang}},
  \bibinfo{author}{\bibfnamefont{Q.}~\bibnamefont{Zhou}},
  \bibinfo{author}{\bibfnamefont{B.}~\bibnamefont{Lian}}, \bibnamefont{and}
  \bibinfo{author}{\bibfnamefont{S.-C.} \bibnamefont{Zhang}},
  \bibinfo{journal}{Phys. Rev. B} \textbf{\bibinfo{volume}{92}},
  \bibinfo{pages}{064520} (\bibinfo{year}{2015}).

\bibitem[{\citenamefont{He et~al.}(2017)\citenamefont{He, Pan, Stern, Burks,
  Che, Yin, Wang, Lian, Zhou, Choi et~al.}}]{He294}
\bibinfo{author}{\bibfnamefont{Q.~L.} \bibnamefont{He}},
  \bibinfo{author}{\bibfnamefont{L.}~\bibnamefont{Pan}},
  \bibinfo{author}{\bibfnamefont{A.~L.} \bibnamefont{Stern}},
  \bibinfo{author}{\bibfnamefont{E.~C.} \bibnamefont{Burks}},
  \bibinfo{author}{\bibfnamefont{X.}~\bibnamefont{Che}},
  \bibinfo{author}{\bibfnamefont{G.}~\bibnamefont{Yin}},
  \bibinfo{author}{\bibfnamefont{J.}~\bibnamefont{Wang}},
  \bibinfo{author}{\bibfnamefont{B.}~\bibnamefont{Lian}},
  \bibinfo{author}{\bibfnamefont{Q.}~\bibnamefont{Zhou}},
  \bibinfo{author}{\bibfnamefont{E.~S.} \bibnamefont{Choi}},
  \bibnamefont{et~al.}, \bibinfo{journal}{Science}
  \textbf{\bibinfo{volume}{357}}, \bibinfo{pages}{294} (\bibinfo{year}{2017}).

\bibitem[{\citenamefont{Ji and Wen}(2018)}]{PhysRevLett.120.107002}
\bibinfo{author}{\bibfnamefont{W.}~\bibnamefont{Ji}} \bibnamefont{and}
  \bibinfo{author}{\bibfnamefont{X.-G.} \bibnamefont{Wen}},
  \bibinfo{journal}{Phys. Rev. Lett.} \textbf{\bibinfo{volume}{120}},
  \bibinfo{pages}{107002} (\bibinfo{year}{2018}).

\end{thebibliography}

\appendix*

\section{Derivation for the number of particles and superconducting gap
equation}

\label{append} In the Hilbert space $\{\Phi^{(n)}\}$, with $\Phi^{(n)}=$
$\left(  \psi_{k,\xi\uparrow}^{(n)},\psi_{-k,\overline{\xi}\downarrow
}^{(n)\ast},\psi_{k,\xi\downarrow}^{(n)},\psi_{-k,\overline{\xi}\uparrow
}^{(n)\ast}\right)  $, the matrix elements of the BdG Hamiltonian are
calculated by
\begin{widetext}%
\begin{align}
\widetilde{\mathcal{H}}_{\mathrm{BdG}}^{(m,n)} &  =\langle\Phi^{(m)}%
|\widetilde{\mathcal{H}}_{\mathrm{BdG}}|\Phi^{(n)}\rangle\nonumber\\
&  =\left(
\begin{array}
[c]{c}%
\psi_{k,\xi\uparrow}^{(m)}\\
\psi_{-k,\overline{\xi}\downarrow}^{(m)\ast}\\
\psi_{k,\xi\downarrow}^{(m)}\\
\psi_{-k,\overline{\xi}\uparrow}^{(m)\ast}%
\end{array}
\right)  ^{\dag}\left(
\begin{array}
[c]{cccc}%
h_{\mathbf{p},\xi\uparrow} & \hat{\Delta}_{2} & 0 & 0\\
\hat{\Delta}_{2}^{\dag} & -h_{-\mathbf{p},-\xi\downarrow}^{\ast} & 0 & 0\\
0 & 0 & h_{\mathbf{p},\xi\downarrow} & \hat{\Delta}_{2}\\
0 & 0 & \hat{\Delta}_{2}^{\dag} & -h_{-\mathbf{p},-\xi\uparrow}^{\ast}%
\end{array}
\right)  \left(
\begin{array}
[c]{c}%
\psi_{k,\xi\uparrow}^{(n)}\\
\psi_{-k,\overline{\xi}\downarrow}^{(n)\ast}\\
\psi_{k,\xi\downarrow}^{(n)}\\
\psi_{-k,\overline{\xi}\uparrow}^{(n)\ast}%
\end{array}
\right)
\end{align}
\end{widetext}
where $\psi_{-k,\xi\sigma}^{(n)\ast}$ is the wavefunction corresponding to the
hole Hamiltonian $h_{-\mathbf{p},\xi\sigma}^{\ast}$. By using the relations
below%
\begin{align}
\langle\psi_{k,\xi\sigma}^{(m)}|h_{\mathbf{p},\xi\sigma}|\psi_{k,\xi\sigma
}^{(n)}\rangle &  =(\Omega_{\xi\sigma}^{n}-\mu_{\xi\sigma})\delta_{m,n},\\
\langle\psi_{-k,\overline{\xi}\overline{\sigma}}^{(m)\ast}|h_{-\mathbf{p}%
,\overline{\xi}\overline{\sigma}}^{\ast}|\psi_{-k,\overline{\xi}%
\overline{\sigma}}^{(n)\ast}\rangle &  =(\mu_{\overline{\xi}\overline{\sigma}%
}-\Omega_{\overline{\xi}\overline{\sigma}}^{n})\delta_{m,n},\\
\langle\psi_{k,\xi\sigma}^{(m)}|\hat{\Delta}_{2}|\psi_{-k,\overline{\xi
}\overline{\sigma}}^{(n)\ast}\rangle &  =\Delta\delta_{m,n},
\end{align}
we can derive the matrix elements of the BdG Hamiltonian to be%
\begin{equation}
\widetilde{\mathcal{H}}_{\mathrm{BdG}}^{(m,n)}=\left(
\begin{array}
[c]{cccc}%
\Lambda_{n,\xi\uparrow} & \Delta & 0 & 0\\
\Delta & -\Lambda_{n,-\xi\downarrow} & 0 & 0\\
0 & 0 & \Lambda_{n,\xi\downarrow} & \Delta\\
0 & 0 & \Delta & -\Lambda_{n,-\xi\uparrow}%
\end{array}
\right)  \delta_{m,n}\label{eq_BdG}%
\end{equation}
with $\Lambda_{n,\xi\sigma}=\Omega_{\xi\sigma}^{n}-\mu_{\xi\sigma}$. Since the
spin index $\sigma$ is locked to the valley index $\xi$, $\Lambda_{n,\xi
\sigma}=\Lambda_{n,\overline{\xi}\overline{\sigma}}$ and, by labeling
$\chi=\xi\sigma$ for brevity, we can express the BdG Hamiltonian as presented
in Eq. (7) of the paper.

To calculate the number of particles, we express the BdG Hamiltonian in the
form of the second quantization as%
\begin{align}
\widetilde{\mathcal{H}}_{\mathrm{BdG}}  &  \mathcal{=}\sum_{\xi\sigma
,n,k}[\Lambda_{n,\xi\sigma}c_{n,k,\xi\sigma}^{\dag}c_{n,k,\xi\sigma}%
-\Lambda_{n,\overline{\xi}\overline{\sigma}}c_{n,-k,\overline{\xi}%
\overline{\sigma}}c_{n,-k,\overline{\xi}\overline{\sigma}}^{\dag}]\nonumber\\
&  +\sum_{\xi\sigma,n,k}(\Delta c_{n,k,\xi\sigma}^{\dag}c_{n,-k,\overline{\xi
}\overline{\sigma}}^{\dag}+h.c.)\ , \label{eq_sh}%
\end{align}
where $c_{n,k,\xi\sigma}^{\dag}$ ($c_{n,k,\xi\sigma}$) creates (annihilates)
an electron in the state $|\psi_{k,\xi\sigma}^{(n)}\rangle$. The number of
particles now can be defined as $\mathcal{N}_{\xi\sigma}=\sum_{nk}\langle
c_{n,k,\xi\sigma}^{\dag}c_{n,k,\xi\sigma}\rangle$. According to the
fluctuation-dissipation theorem~\cite{1367-2630-18-9-093040}, i.e.,
\begin{equation}
\langle c_{n,k,\xi\sigma}^{\dag}c_{n,k,\xi\sigma}\rangle=-\int\frac{d\omega
}{\pi}f(\omega)\mbox{Im}\langle\langle c_{n,k,\xi\sigma}|c_{n,k,\xi\sigma
}^{\dag}\rangle\rangle_{\omega}^{r}\ ,
\end{equation}
where $\langle\langle c_{n,k,\xi\sigma}|c_{n,k,\xi\sigma}^{\dag}\rangle
\rangle_{\omega}^{r}$ represents the retarded Green's function and
$f(\omega)=1/(1+e^{\omega/k_{B}T})$ is the Fermi-Dirac distribution function,
the number of particles can be expressed as%
\begin{equation}
\mathcal{N}_{\chi}=-gN_{\phi}\sum_{n}\int_{-\infty}^{\infty}d\omega
\operatorname{Im}[G_{\chi,11}^{r}(\omega)/\pi]f(\omega)\ ,
\end{equation}
where $g=2$ is the valley$\times$spin degeneracy and $N_{\phi}=A/(2\pi
l_{B}^{2})$ is degeneracy of the LL originating from the summation over $k$,
with $A=L_{x}L_{y}$ as area of the sample. The retarded Green's function, with
respect to $\mathcal{H}_{\chi}$, is defined as%
\begin{align}
G_{\chi}^{r}(\omega)  &  =\frac{1}{\omega+i0^{+}-\mathcal{H}_{\chi}%
}\nonumber\\
&  =\frac{1}{2}\sum_{\eta=\pm}\frac{1}{\omega^{+}-\eta E_{n}^{\chi}}\left[
\tau_{0}+\frac{1}{\eta E_{n}^{\chi}}\left(
\begin{array}
[c]{cc}%
\varepsilon_{n,\chi} & \Delta\\
\Delta & -\varepsilon_{n,\chi}%
\end{array}
\right)  \right]  \label{eq_retard}%
\end{align}
with $E_{n}^{\chi}=\sqrt{\varepsilon_{n,\chi}^{2}+\Delta^{2}}$ and
$\varepsilon_{n,\chi}=\Omega_{\chi}^{n}-\mu_{\chi}$, where we adopted the
abbreviation $\omega^{+}=\omega+i0^{+}$. Therefore, we can obtain for%
\begin{align}
\mathcal{N}_{\chi}  &  =\frac{gN_{\phi}}{2}\sum_{n}\{1-\frac{\varepsilon
_{n,\chi}}{E_{n}^{\chi}}[f(-E_{n}^{\chi})-f(E_{n}^{\chi})]\}\nonumber\\
&  =\frac{gN_{\phi}}{2}\sum_{n}\{1-\frac{\varepsilon_{n,\chi}\tanh[E_{n}%
^{\chi}/(2k_{B}T)]}{E_{n}^{\chi}}\}. \label{eq_Nchi}%
\end{align}
If the deep-energy states $n<n_{\mathrm{F}}$ are fully occupied, with
$n_{\mathrm{F}}$ labeling the highest occupied LL, we can reduce
Eq.\ (\ref{eq_Nchi}) to Eq. (\ref{slef_Nx}) in the paper. In the low
temperature and weak interaction limit $T,\Delta\ll\hbar\omega_{c}$, we can
approximate $\tanh[\sqrt{\varepsilon_{n,\chi}^{2}+\Delta^{2}}/(2k_{B}%
T)]\simeq1$ for $n<n_{\mathrm{F}}$ and $\sqrt{\varepsilon_{n_{\mathrm{F}}%
,\chi}^{2}+\Delta^{2}}\simeq|\varepsilon_{n_{\mathrm{F}},\chi}|$. Therefore,
we can further reduce Eq. (\ref{slef_Nx}) to Eq. (\ref{filfactor}). For
$k_{B}T\ll\Delta$, $\tanh[E_{n_{\mathrm{F}}}^{\chi}/(2k_{B}T)]\simeq1$ and
from Eq. (\ref{filfactor}), it is easy to solve the chemical potential, as
given by Eq.\ (\ref{zero_mu}).

The gap equation, similar to Eq. (\ref{FDT_N}), is obtained self-consistently
by the mean-field approximation, i.e.,%
\begin{align}
\Delta(T,\nu_{\chi})  &  =Ug\frac{1}{A}\sum_{nk}\langle c_{n,-k,\overline{\xi
}\overline{\sigma}}^{\dag}c_{n,k,\xi\sigma}^{\dag}\rangle\nonumber\\
&  =-Ug\overline{N}_{\phi}\sum_{n}\int\frac{d\omega}{\pi}f(\omega
)\mbox{Im}\langle\langle c_{n,k,\xi\sigma}^{\dag}|c_{n,-k,\overline{\xi
}\overline{\sigma}}^{\dag}\rangle\rangle_{\omega}^{r}\nonumber\\
&  =-Ug\overline{N}_{\phi}\sum_{n}\int_{-\infty}^{\infty}\frac{d\omega}{\pi
}\mbox{Im}[G_{\chi,12}^{r}(\omega)]f(\omega) \label{eq_delta}%
\end{align}
with $\overline{N}_{\phi}=N_{\phi}/A$ as the number of flux quanta per unit
area. By substituting the retarded Green's function, Eq.\ (\ref{eq_retard}),
into the above equation, we can derive Eq. (\ref{self_gap}).
\end{document}